\documentclass{iopart}

\expandafter\let\csname equation*\endcsname\relax
\expandafter\let\csname endequation*\endcsname\relax

\usepackage[acronym,shortcuts,sanitize=none]{glossaries}
\usepackage[colorlinks,linkcolor={blue},citecolor={blue},urlcolor={red}]{hyperref}
\usepackage{iopams}
\usepackage{subfig}
\usepackage[margin=10pt,font=small,format=hang]{caption}
\usepackage{multicol}
\usepackage{multirow}
\usepackage{array}
\usepackage{graphicx}
\usepackage{xcolor}
\usepackage{cleveref}

\DeclareGraphicsExtensions{%
    .pdf,.PDF,%
    .png,.PNG,%
    .jpg,.mps,.jpeg,.jbig2,.jb2,.JPG,.JPEG,.JBIG2,.JB2}

\begin{document}

\title{Improving the Data Quality of Advanced LIGO Based on Early Engineering Run Results}
\author{
L.~K.~Nuttall$^{1}$, T.~J.~Massinger$^{1}$, J.~Areeda$^{2}$, 
J.~Betzwieser$^{3}$, S.~Dwyer$^{4}$, A.~Effler$^{5}$, R.~P.~Fisher$^{1}$, 
P.~Fritschel$^{6}$, J.~S.~Kissel$^{4}$, A.~P.~Lundgren$^{7}$, 
D.~M.~Macleod$^{8}$, D.~Martynov$^{5,6}$, J.~McIver$^{5,9}$, 
A.~Mullavey$^{3}$, D.~Sigg$^{4}$, J.~R.~Smith$^{2}$, 
G.~Vajente$^{5}$, A.~R.~Williamson$^{10}$, C.~C.~Wipf$^{5}$
}
\address{$^{1}$Syracuse University, Syracuse, NY 13244, USA}
\address{$^{2}$California State University Fullerton, Fullerton, CA 92831, USA}
\address{$^{3}$LIGO Livingston Observatory, Livingston, LA 70754, USA}
\address{$^{4}$LIGO Hanford Observatory, Richland, WA 99352, USA}
\address{$^{5}$LIGO, California Insitute of Technology, Pasadena, CA 91125, USA}
\address{$^{6}$LIGO, Massachusetts Institute of Technology, Cambridge, MA 02139, USA}
\address{$^{7}$Data Analysis Group, Albert-Einstein-Institut, Max-Planck-Institut f\"{u}r Gravitationsphysik, D-30167 Hannover, Germany}
\address{$^{8}$Louisiana State University, Baton Rouge, LA 70803, USA}
\address{$^{9}$University of Massachusetts-Amherst, Amherst, MA 01003, USA}
\address{$^{10}$Cardiff University, Cardiff, CF24 3AA, UK}

\ead{laura.nuttall@ligo.org}
 
\begin{abstract}
The Advanced Laser Interferometer Gravitational-wave Observatory (LIGO) 
detectors have completed their initial upgrade phase and will enter the first 
observing run in late 2015, with detector sensitivity expected to 
improve in future runs. Through the combined efforts 
of on-site commissioners and the Detector Characterization group of the LIGO 
Scientific Collaboration, interferometer performance, in terms of data quality, 
at both LIGO observatories has vastly improved from the start of commissioning 
efforts to present. Advanced LIGO has already surpassed Enhanced LIGO in 
sensitivity, and the rate of noise transients, which would negatively impact 
astrophysical searches, has improved. Here we give details of some of the work 
which has taken place to better the quality of the LIGO data ahead of the first 
observing run.
\end{abstract}
\pacs{04.80.Nn.} %GW detectors and experiments
\maketitle

% introduction and low latency pipelines
\section{Introduction}\label{sec:intro}
The Laser Interferometer Gravitational-wave Observatory (LIGO), 
comprised of two 4-kilometer laser interferometers in Hanford, WA (LHO) 
and Livingston, LA (LLO), has been in an upgrade phase since 2010 to bring 
about the second generation of gravitational-wave detectors. Prior to this, 
LIGO was operated in two iterations, named Initial and Enhanced LIGO. The 
former data taking period spanned $2002 - 2007$ in five separate science runs; 
the fifth science run (November 2005 - September 2007) achieved design 
sensitivity in a continuous data-taking fashion \cite{Abbott:2009ab}. 
The laser source and readout system were then upgraded \cite{Fricke:2012dc}, 
with Enhanced LIGO (science run six (S6): July 2009 - October 2010) reaching 
a root-mean-square strain sensitivity of $2\times10^{-22}$ in its most 
sensitive frequency region ($\sim$100 Hz), approximately 30\% more 
sensitive than Initial LIGO \cite{Aasi:2015fr}.

The Advanced LIGO (aLIGO) project has completely upgraded the interferometers 
(this will be discussed in further detail in Section \ref{sec:aLIGO} 
\cite{Aasi:2015fr}), which at design sensitivity will yield a factor of 10 
increase in strain 
sensitivity. In terms of distance, aLIGO expects to be capable of detecting 
the merger of a binary neutron star system, averaged over sky location and 
orientation, to $\sim$200 Mpc \cite{Aasi:2013pr}.
%This is also known as the inspiral detection range.
When achieving this detection range, aLIGO is likely to detect the merger of 
tens of compact binary coalescences (CBCs) every year \cite{Abadie:2010ra}.

Construction of Advanced LIGO began in April 2008; early 
2011 saw the first new hardware being installed. The Livingston 
detector was completed in mid-2014, with Hanford finishing later in the same 
year \cite{Aasi:2015fr}. Both detectors have since achieved lock stretches in 
excess of two hours, with all of the detector's core control systems engaged.
%thus completing their initial requirements. 
Commissioners have since been optimizing the operation of the detectors and 
hunting sources of noise to improve sensitivity ahead of the first observing 
run (scheduled for Fall 2015).

Gravitational wave interferometer data are typically non-stationary; there 
are many long and short duration artifacts in the data. Long duration 
continuous wave searches and gravitational wave background searches are most 
affected by elevated noise at a given frequency, such as variations in 
spectral line amplitudes. This reduces the ability to search over the data at 
these given frequencies. However, transient gravitational wave searches 
(including CBCs and gravitational wave bursts) are most sensitive to short 
duration noise events or `glitches'. These can occur for a variety of reasons, 
including environmental or instrumental mechanisms, some of which are not fully 
understood. 
Utilizing knowledge of the expected gravitational waveform, signal-based 
methods are used by CBC search pipelines to distinguish between noise and a 
gravitational wave signal \cite{Klimenko:2008cw, Sutton:2010xp, Harry:2011co, 
Babak:2013ih}. The search pipelines also rely on studies of the behaviour 
of the detector to accurately remove or `veto' data which are likely to contain 
noise artifacts. Such studies during LIGO's sixth science run, and the 
improvements made to gravitational wave analyses, can be found in 
\cite{Aasi:2015de}.

This paper describes efforts made to characterize the aLIGO detectors before 
the first observing run. Engineering runs are stretches of time where the 
detectors are operated as 
though in an observing run. Several runs have been performed to date, each 
scheduled at various stages of installation or detector configuration. 
The work presented focusses around the sixth engineering run 
(ER6: 8th - 17th December 2014), where only the Livingston detector 
participated. Many of the cases presented were found in the Livingston data, 
however similar artifacts were found in the Hanford data at a later date and 
were easily mitigated due to this early work.

The goal of this work is to identify sources of noise 
which would limit gravitational wave searches, track down their causes, and 
fix the issues at the sources (i.e. the detector). This will improve 
the quality 
of the data, reduce the amount of time needed to be vetoed from an 
analysis, and improve the searches performed. This work will also reduce the 
likelihood that a real signal will be rendered `suspicious' by looking like a 
known glitch.

Section \ref{sec:aLIGO} discusses the Advanced LIGO detectors, including the 
upgrades which have been performed from initial LIGO. Section \ref{sec:methods} 
details some of the tools and methods used to identify 
noise sources, with Section \ref{sec:DQ} describing 
examples which have been identified and mitigated. Section \ref{sec:comparison} 
discusses the output of the detectors, comparing early commissioned aLIGO data 
and data taken in S6 to more recent, mature aLIGO data. A short 
conclusion is given in Section \ref{sec:conclusion}.

\section{The Advanced LIGO Interferometers}\label{sec:aLIGO}
\begin{figure}
\centering
\includegraphics[width=\textwidth]{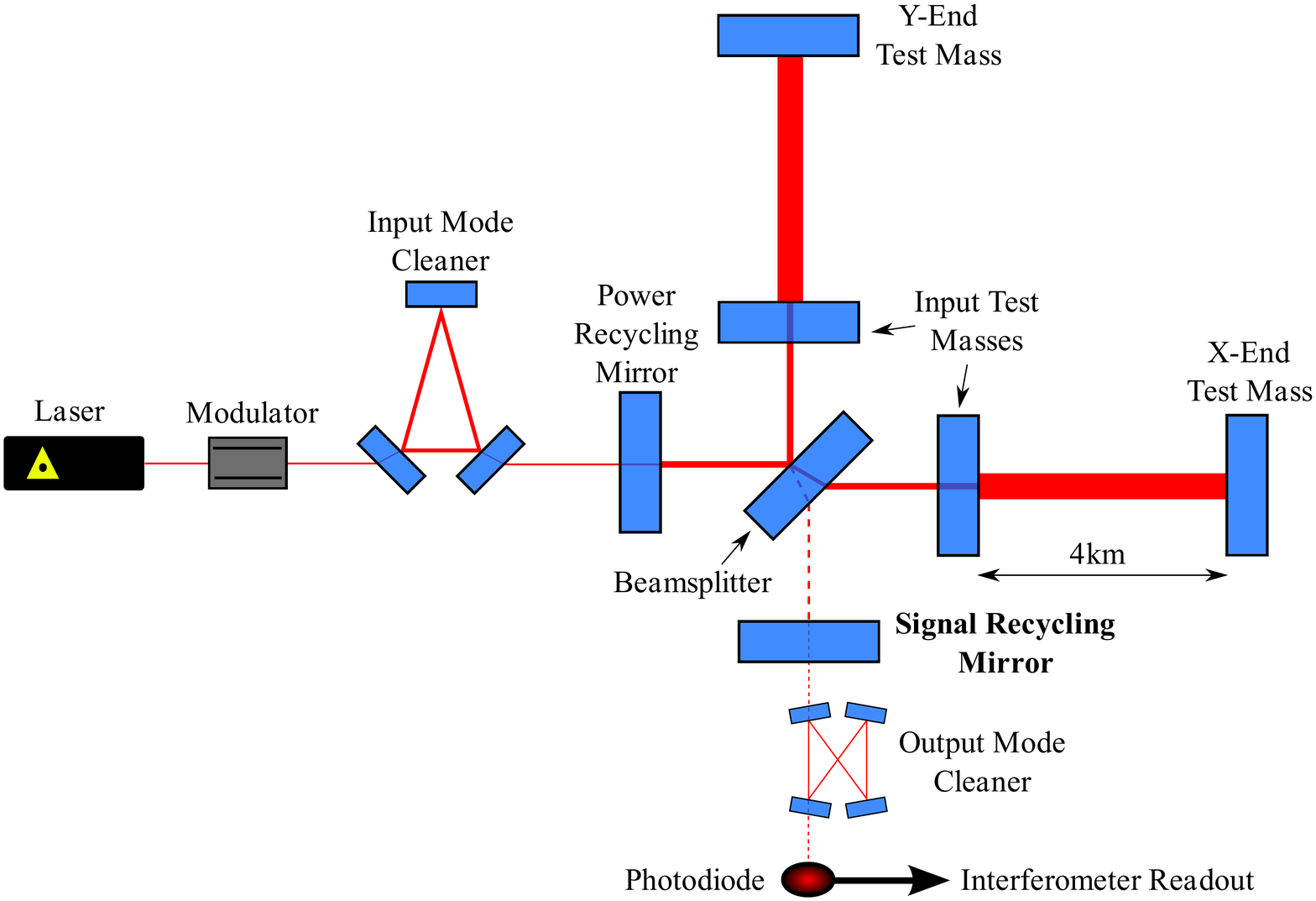}
\caption{\label{aligo_layout}A simplified layout of the Advanced LIGO 
interferometers taken from \cite{Smith:2009al}. The signal recycling mirror is 
the main change in optical layout from Enhanced LIGO.}
\end{figure}

The Advanced LIGO instruments are a pair of modified Michelson interferometers
that employ Fabry-Perot cavities in their arms to increase the interaction time
with a gravitational wave signal \cite{Aasi:2015fr}. A simplified schematic of 
the aLIGO interferometers can be seen in Figure \ref{aligo_layout} 
\cite{Smith:2009al}. The layout of the advanced interferometers is similar in 
some respects to their Enhanced counterparts (for more details of the 
Enhanced detectors, see \cite{Fricke:2012dc, Smith:2009al, 
Smith-Lefebvre:2011om}).

Advanced LIGO interferometers rely on the performance of a series of 
interconnected subsystems to maintain stable operation. The layout of the 
interferometer starts with a Nd:YAG laser that generates a carrier beam at 
1064 nm, \cite{Kwee:2012kw} which is then passed through an electro-optic 
modulator. 
Here, radio-frequency (RF) sidebands are added, which are used for sensing 
and control of the test mass suspensions. The input mode cleaner (IMC), a 
triangular optical cavity, stabilizes the frequency and spatial mode content of 
the beam entering the heart of the interferometer. The effective laser power in 
the arms is increased using a power recycling cavity, which helps to improve 
sensitivity at higher frequencies where the detectors are limited by shot 
noise. New to Advanced LIGO is signal recycling, which allows the 
interferometer to be operated with a broader frequency response 
\cite{Meers:1988sr}. Using improved seismic isolation, test mass 
suspensions and heavier test masses \cite{Aasi:2015fr}, the low frequency limit 
on the useful sensitivity where LIGO performs an astrophysical search is pushed 
down from 40 Hz to 10 Hz. At higher frequencies, higher input laser power and 
better optical coatings \cite{Harry:2006th, Harry:2007th} are responsible for 
increased sensitivity. 

Detection of gravitational waves in aLIGO interferometers is done by measuring
the phase mismatch incurred when the two arms experience differential motion.
The degree of freedom describing the differential length of the arms is labeled
`DARM'. The instrument is considered `locked' when
%there is appreciable power circulating in the arms and 
all cavity lengths are held within their linear 
operating range. The differential arm length is controlled by a 
feedback loop and the error signal of the DARM feedback loop is calibrated to 
measure the external strain affecting the cavities. This signal is used to 
perform the astrophysical searches and, as such, is the main subject of data 
quality investigations.

There are a large number of recorded signals used to monitor and control the
interferometers that are not directly used in the astrophysical searches, but 
are still extremely useful for detector characterization. These channels are
referred to as auxiliary channels and can be used to track down the causes
of systematic noise in DARM. Examples of these auxiliary channels include 
environmental monitors and signals used to control the auxiliary
degrees of freedom in the instrument. These channels are used very often
in detector characterization and are heavily featured in Section 
\ref{sec:DQ}.

\section{Detector Characterization Methods}\label{sec:methods}
The Detector Characterization\footnote{A data analysis group within the LIGO 
Scientific Collaboration (LSC) working to characterize the performance of the 
LIGO instruments, and quantify data quality as used by the search working 
groups \cite{Cavaglia:2015or}.} group employs a diverse set of methods 
to assess data quality and conduct investigations when the need arises. 
The goal of these investigations, in general, is to ensure that a 
gravitational wave detection can be made should a candidate event occur. 
This being the case, the definition of data quality is highly dependent on 
how a feature in the output of the instrument is witnessed and processed by 
a gravitational wave search pipeline. The ideal outcome of a data quality 
investigation is to fix the problem at the source so that the data analyzed is 
clean of artifacts. The alternative outcome is to find a way to 
systematically identify the noise so that it can be excised from the 
analysis time in the form of a veto.

\subsection{Starting a Data Quality Investigation}
There are two common ways for a data quality investigation to arise: an unusual 
feature in the raw instrumental data is witnessed (such as in a spectrum or 
spectrogram) or an unusual feature in the output of a gravitational wave search 
pipeline is found. Unusual features are typically large, visible to the eye, 
features that are not coincident between detectors. Subsequent investigations 
aim to provide a complete 
description of instrumental features and how they influence the output of a 
gravitational wave search. 
%In the former case, an issue is often found by 
%on-site commissioners or by members of the Detector Characterization group, 
%that are monitoring the data quality both on and off site.

A convenient starting point for investigating troublesome instrumental 
features is the LIGO summary pages \cite{Macleod:2014lh, Macleod:2014ll}. These 
webpages house and archive the state and the behavior of the interferometer 
from the perspective of each subsystem with a focus on assessing data quality. 
There are approximately ten subsystems in total, which include, for example, 
the pre-stabilized laser (PSL), input mode cleaner (IMC), and alignment sensing 
and control (ASC) subsystems. 
It is common to begin an instrumental investigation by searching through the 
summary pages and checking each subsystem page for any clues that might 
identify the source of a feature.

Changes or irregularities in the DARM spectrum are more easily 
identified and resolved using a suite of information that includes the 
typical behavior of the instrument, the results of prior Detector 
Characterization studies, and any configuration changes noted by 
commissioners in the site logbook. 

\subsection{Data Quality Tools}
A variety of tools are used to identify and pinpoint sources of noise. 
These include tools designed to represent the output of a typical gravitational 
wave search to highlight poor data quality times. Using the output of a search 
requires a careful understanding of the behavior of each gravitational wave 
search pipeline, so that signal processing artifacts are not confused for 
problems in the instrument. There are also tools which search for 
auxiliary channels that are coherent with DARM or determine the statistical 
significance of glitches found in both auxiliary channels and DARM.

Omicron \cite{Robinet:2014om}, a transient search algorithm, gives a reliable 
account of a typical transient gravitational wave search. It is derived from an 
established transient search pipeline called the Q-pipeline or Omega 
\cite{Chattergi:2005om}, 
which is a transient search algorithm that produces triggers 
based on a sine-Gaussian excess power method. For each trigger a central time, 
central frequency, duration, bandwidth, Q-value and signal-to-noise ratio (SNR) 
(or normalised tile energy) is provided. Omicron is sensitive to most glitches 
and reports useful information about the properties of a glitch; it is run in 
quasi-real time over a large number of detector channels.

As well as being useful for assessing the data quality of a transient search, 
Omicron triggers can also be fed into other tools which search for interesting 
instrumental correlations. For example, Hierarchical Veto (HVeto) 
\cite{Smith:2011hv} 
reads in a set of Omicron triggers produced from DARM, and compares them to 
Omicron triggers produced from auxiliary channels. If there are glitches in 
auxiliary channels that are coincident in time with glitches in DARM to an 
extent that is statistically interesting, an auxiliary channel is marked as 
having high significance and warrants further investigation. However there are 
some auxiliary channels with known couplings to DARM. To thoroughly assess 
whether a channel directly leads to DARM, loud signals are injected in to the 
interferometer so that two lists of auxiliary channels can be compiled, one 
which is `safe' and another `unsafe'. 
The unsafe channel list details the auxiliary 
channels with a known coupling to DARM and the safe channel list those without. 
It is only the safe channels which are used by tools like HVeto. 
Since these auxiliary channels have negligible sensitivity to gravitational 
waves, and therefore should not witness signals originating in the DARM degree 
of freedom, any statistical correlation points to an auxiliary degree of 
freedom with some unintended noise coupling with DARM. Examples of this 
procedure will be presented in Section \ref{sec:DQ}.

Advanced LIGO interferometers rely on the performance of a series of 
interconnected 
subsystems to maintain stable operation. Each subsystem has a team responsible
for its maintenance and commissioning. In an effort to better understand the
output of the instrument and its impact on astrophysical search pipelines,
the Detector Characterization group has mirrored this approach and assigned 
data quality liaisons to each subsystem. These subsystem liaisons are tasked 
with understanding the operation of their respective subsystems and are 
responsible for interfacing with commissioners, developing summary pages that 
allow those interested in data quality to have a general overview of the 
subsystem, and 
designing real-time monitoring of their subsystem's data quality using the 
Online Detector Characterization (ODC) framework \cite{Ballmer:2013od}.

\section{Data Quality Issues}\label{sec:DQ}
During S6 there were numerous examples of sources of noise identified in the 
interferometers, which could not be fixed immediately or at all during the 
science run \cite{Aasi:2015de}. `Data Quality (DQ) flags' were constructed to 
identify noisy times as they would adversely affect a gravitational wave 
search. Analysis groups would use these flags to make an informed decision on 
which data to analyse. Teams of people, both on and off site, have been working 
to identify and fix sources of noise which would impact a search for 
gravitational waves before the run begins. This will hopefully have the 
added bonus of minimizing the number of DQ flags which will need to be used 
by the analyses in the first observing run. The remainder of this section 
highlights some of the work 
done at both LHO and LLO to mitigate noise prior to the observing run, however 
focus is given to studies at LLO since this observatory has been in active 
commissioning longer than LHO.

\subsection{DAC Calibration Glitches}
In the commissioning period leading up to the 6th Engineering Run, a 
population of low frequency 
(10 Hz - 100 Hz) glitches appeared in the output of the instrument. 
While monitoring the locked instrument, it was noticed that these 
glitches were correlated with times when certain suspension control signals 
crossed a value of zero. 

The aLIGO suspension system uses 18-bit digital-to-analog converters 
(DACs) to interact with various systems within the interferometers. These 
18-bit DACS are composed of a 16-bit low-order DAC and a 2-bit high-order DAC. 
%The 2-bit DAC has a weighting factor which is four times that of the 16-bit 
%DAC. The four states of the 2-bit DAC are individually calibrated during 
%autocalibration, however during operation they are corrected in real-time. 
%The calibration of the 16-bit DAC is such that it occupies exactly one state of the 2-bit DAC.
The 2-bit chip has four states which are individually calibrated (during 
an auto-calibration routine), and then are corrected in real-time. 
The calibration of the 16-bit chip is such that is occupies exactly one state 
of the 2-bit chip \cite{General:2010gs}. 

\begin{figure}
\centering
\includegraphics[width=\textwidth]{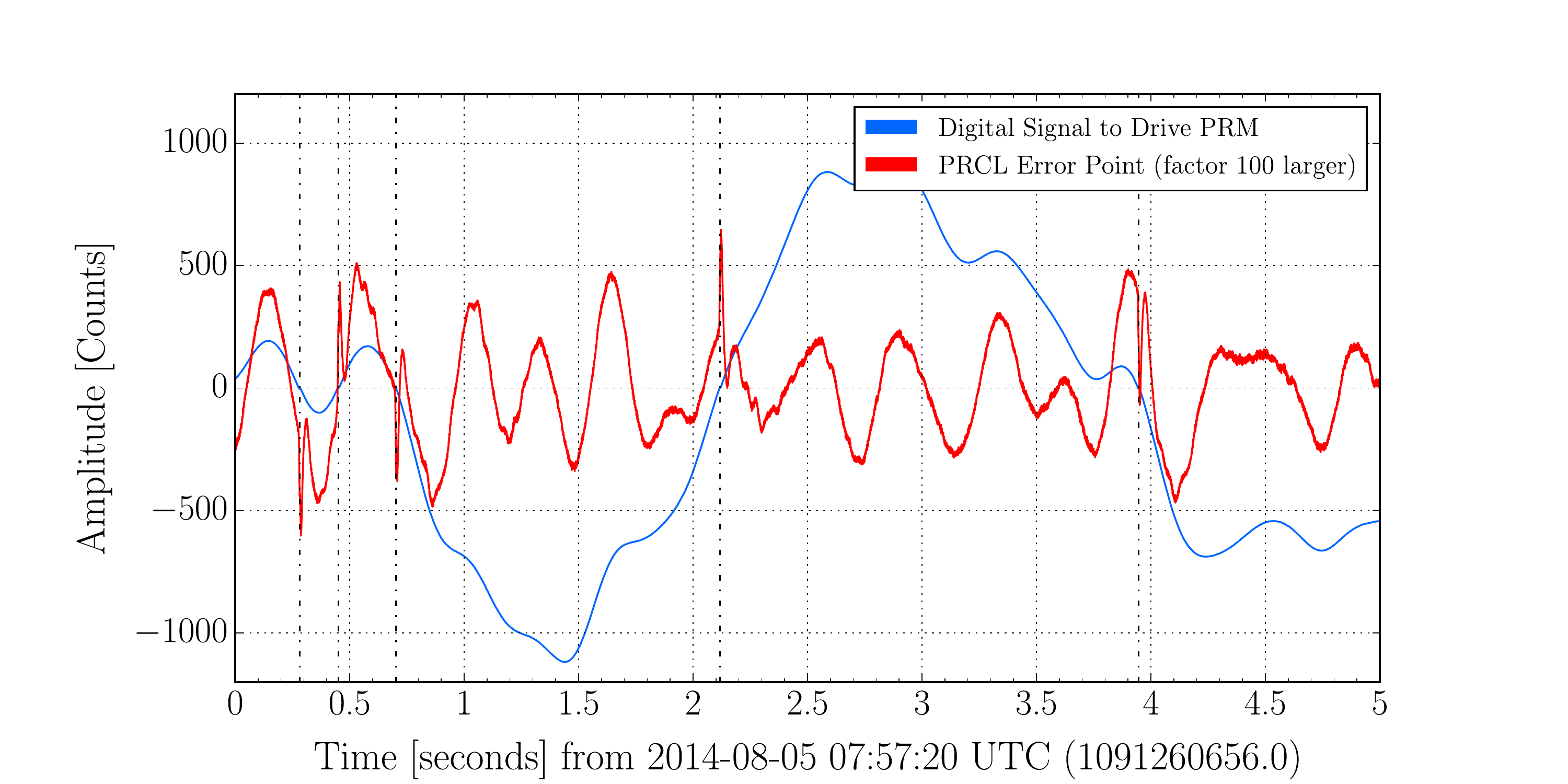}
\caption{\label{dac_glitches}Figure \ref{dac_glitches} is a timeseries of the 
drive signal sent to the DAC for the last stage of the power recycling mirror 
(PRM) suspensions along with the power recycling cavity length (PRCL) error 
point signal, multiplied by a factor of 100 for scaling purposes. Everytime the PRM drive signal crosses a value of zero a discontinuity is seen in the error 
point signal. There are 5 such examples in this plot.}
\end{figure}

In these 18-bit DACs glitches have been observed at both sites due to an output 
discontinuity when the 2-bit DAC engages. 
This discontinuity is caused by a voltage calibration error between the 16-bit
and 2-bit chips in the DAC. In these DACs, the 2-bit chip is responsible for 
the two highest order bits, which causes the discontinuity to occur for 
transitions at output values of zero and $\pm2^{16}$ 
(Figure \ref{dac_glitches}).
%These zero and $\pm2^{16}$ crossings of the suspension DACs caused short, 
%loud glitches in the interferometer DARM channel between $\sim$10-100 Hz. 
These glitches were particularly detrimental to transient search pipelines, as 
they increase the rate of loud, low frequency background events, making the 
ability to confidently identify a true signal difficult.

To fully understand the scope of the problem and its impact on astrophysical 
searches, software was developed that 
systematically searched through suspension drive signals for any instances 
when they crossed either zero or $\pm2^{16}$. Any time this happened, the time 
and offending suspension drive signal channel were recorded and fed in to 
Hveto, which was able to discover which channels had crossings that were 
statistically correlated with glitches in DARM. This served two purposes: it 
provided the capability to pinpoint which suspensions 
required immediate intervention to clean up the output of the instrument, and 
it quantitatively demonstrated the effect of the DAC calibration glitches on 
data quality for a transient search.

Two strategies were employed to try to mitigate this type of glitch. First, an 
offset was applied to the signals so they would never cross values of zero or 
$\pm2^{16}$. However, this did not fix the root problem and presented an easy 
opportunity for the glitches to resurface should the signals begin to drift. 
Offsetting the output signals to avoid these values also 
limit the dynamic range of the control signals being sent to the 
optics, which had the potential to increase the influence of noise associated 
with the DACs. The solution, instead, was found to be an auto-calibration 
routine. In this routine an internal voltage reference is used to calibrate 
the two chips such that when the 2-bit DAC engages a smooth transition occurs 
(rather than a step function). Re-calibration of the 18-bit DACs caused the 
discontinuity to nearly 
vanish \cite{Betzwieser:2014da}. It was observed, however, that over the 
course of a few weeks after the calibration was performed, the calibration 
would degrade and the discontinuity would return. Regular calibration of the 
18-bit DACs in the aLIGO suspension system is required and has been 
implemented to minimize these glitches. More stable solutions are under study.
The software mentioned previously is currently employed to continuously 
monitor the status of the DAC calibration and determine its correlation to 
glitches in DARM.

\subsection{Radio Frequency Beat Notes}
\begin{figure}
\centering
\subfloat[Example from LHO\label{whistle_lho}]{\includegraphics[width=\textwidth]{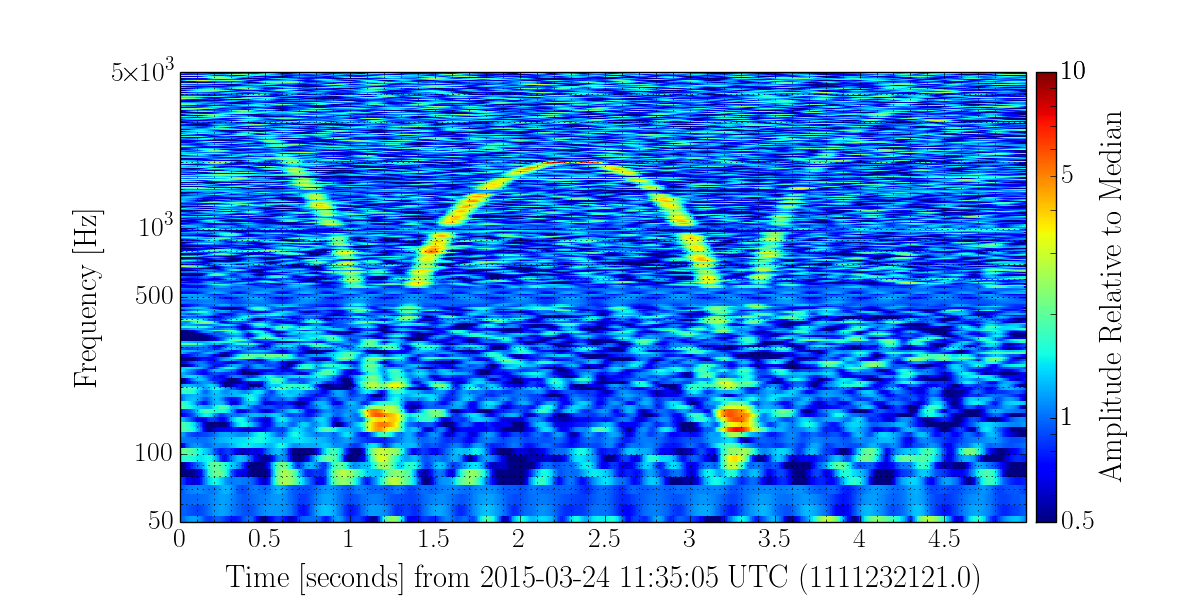}}\\
\subfloat[Example from LLO\label{whistle_llo}]{\includegraphics[width=\textwidth]{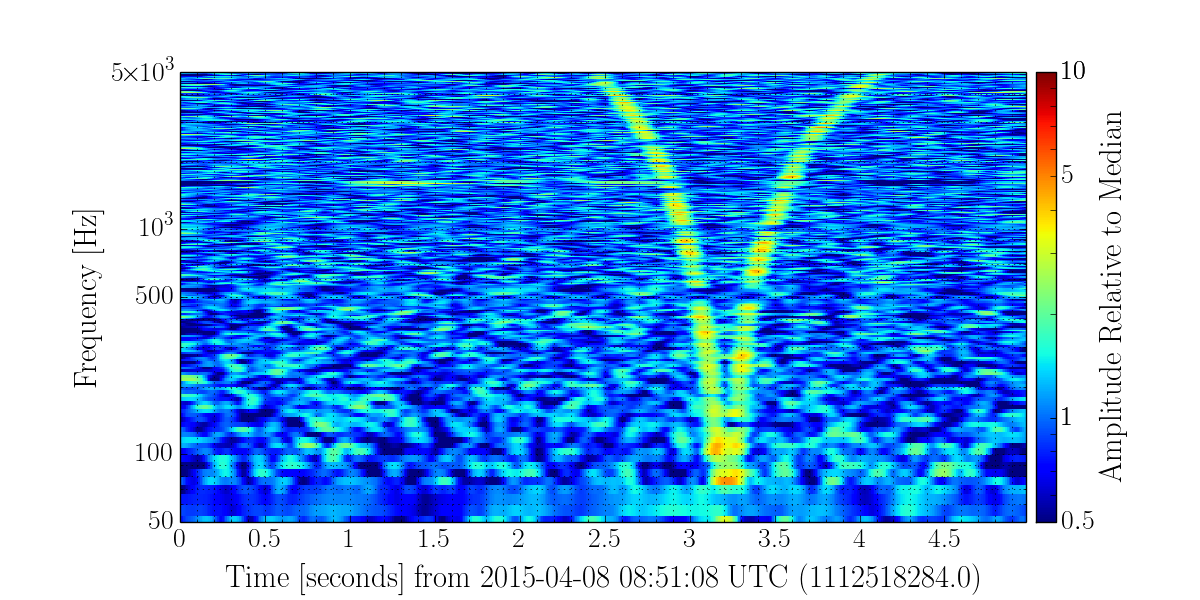}}
\caption{\label{arches_both}Radio Frequency beat notes seen in the DARM channel 
at both Hanford and Livingston are shown in Figures \ref{whistle_lho} and 
\ref{whistle_llo} respectively. Both plots are normalised spectrograms, 
illustrating examples of the characteristic `W' and `V' shapes, produced when 
two oscillators sweep past one another in frequency space.}
\end{figure}

\begin{figure}
\centering
\includegraphics[width=\textwidth]{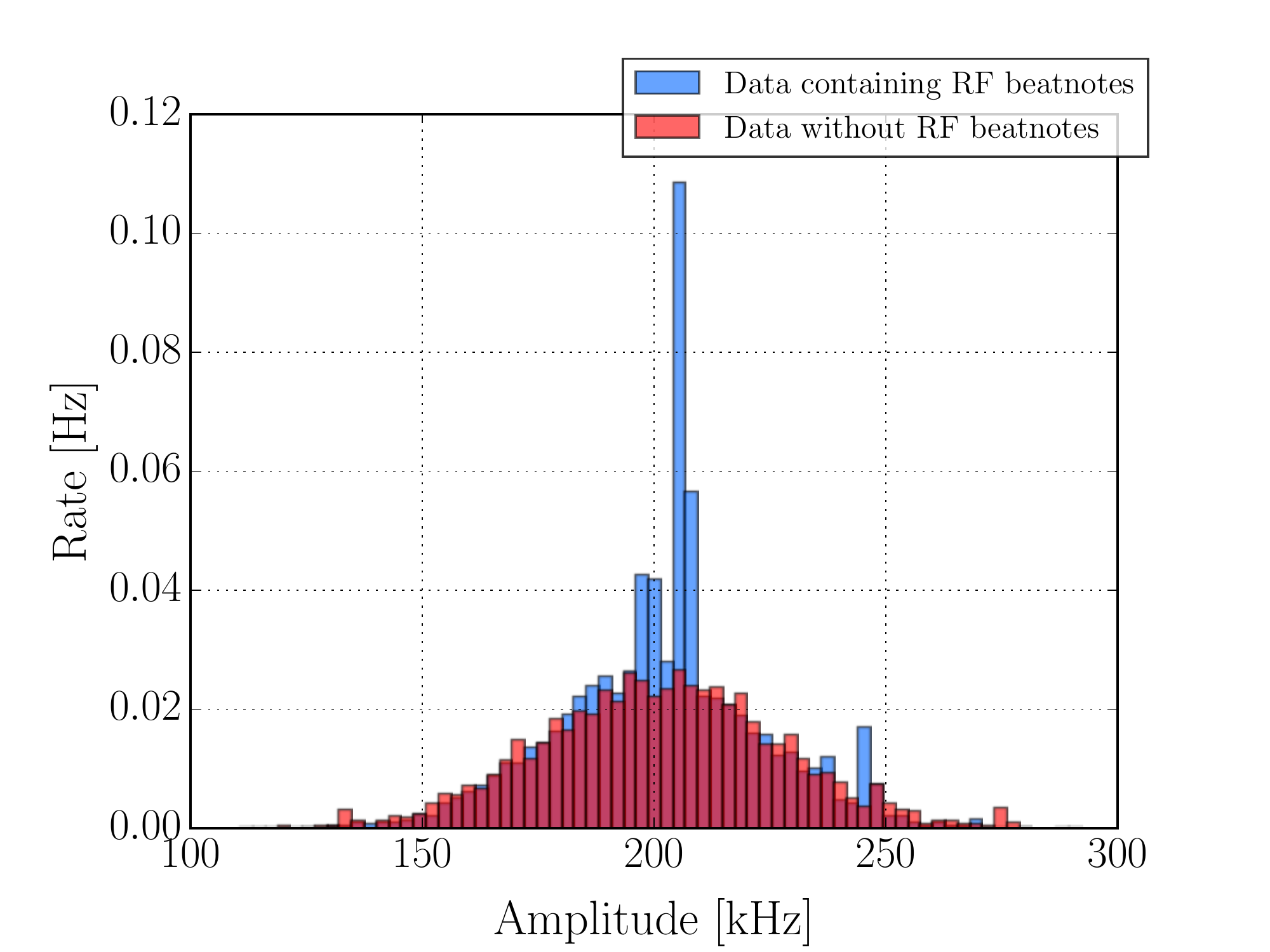}
\caption{\label{imc_hist}A rate histogram plot of the Omicron DARM triggers 
over $\sim$six hour period. Plotted on the x-axis is the value of a 
particular channel at every time when a trigger was produced.
This channel was found to be a good witness of the beat notes. 
At values of $\sim$ 196-201, 204-209 and 245-250 kHz beat 
notes were observed in DARM (blue bars). Without beat notes the distribution 
of triggers appears Gaussian (red bars).}
\end{figure}

Throughout aLIGO commissioning, radio frequency beat notes in auxiliary 
channels have been observed and, as the noise in the instruments improved, 
signals caused by radio frequency beat notes became prevalent in DARM. Across a 
wide range of frequencies, signals were observed in DARM spectrograms with a 
characteristic `W' or `V' shape as shown in Figure \ref{arches_both}.
Initial investigations pointed towards a channel that monitors the detuning of 
the laser frequency based on the input mode cleaner length as being a good 
predictor of this feature. Every time 
this channel had particular values these beat notes would be observed in DARM, as can be seen 
in Figure \ref{imc_hist}. This figure shows a rate histogram of DARM triggers 
(as seen by Omicron) over $\sim$six hour period binned by the corresponding 
channel value (blue bars). When the channel had 
values of $\sim$ 196-201, 204-209 and 245-250 kHz (according to Figure 
\ref{imc_hist}), beat notes were observed in DARM. In the absence of beat 
notes, the distribution of the triggers appears Gaussian (red bars).

Voltage controlled oscillators (VCOs) are used in control loops throughout
the LIGO interferometers. For example, a voltage controlled oscillator in the
pre-stabilized laser frequency stabilization servo (FSS) generates a radio
frequency signal which drives an acousto-optic modulator
(AOM). The AOM generates frequency sidebands on the laser which are used as a 
frequency reference in the FSS loop.
This loop is used to correct frequency fluctuations in the laser by locking
the frequency of the laser to the length of the input mode cleaner
\cite{Angert:2009vc}. Beat notes between voltage controlled oscillators are
visible in the stored detector output when the absolute frequency difference
between two voltage controlled oscillators drops below 16 kHz. 

\begin{figure}
\centering
\includegraphics[width=\textwidth]{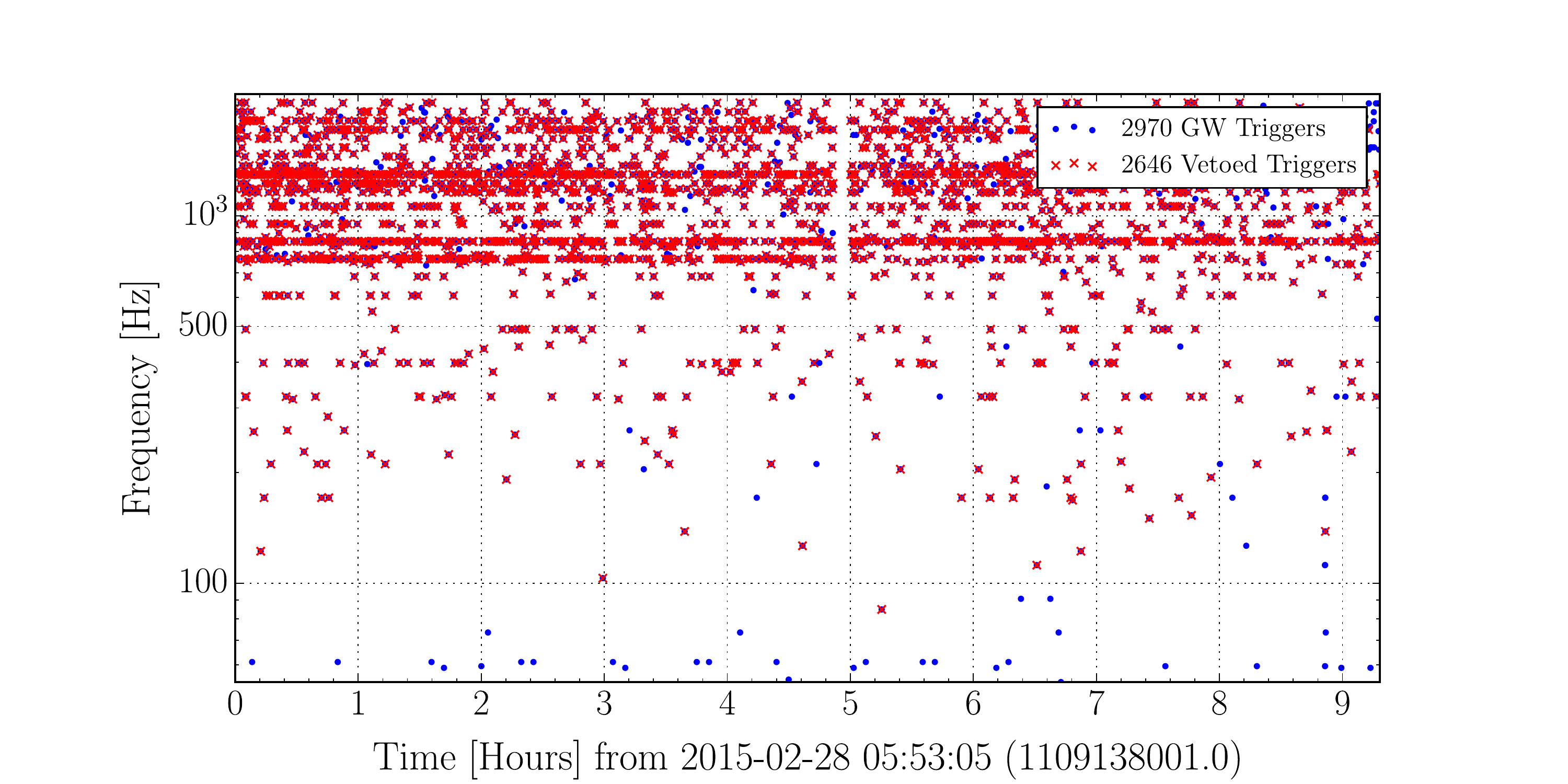}
\caption{\label{hveto_llo}Noise events in the DARM data recorded by Omicron 
over $\sim$nine hour time period. The blue points represent all DARM 
triggers for this period and the red `x' overlay indicates those identified by 
HVeto as being coincident with an auxiliary channel which was witnessing the 
radio frequency beat notes. Approximately 90\% of the DARM triggers are vetoed 
by Hveto.}
\end{figure}

Despite efforts 
to keep radio frequency radiation and pick up to a minimum, beat notes can 
still contaminate the low noise DARM channel. It was found that these 
features occurred every time the frequency stabilization servo voltage 
controlled oscillator at LLO or LHO swept through 79.2 MHz, the same frequency 
as a fixed frequency fiber acousto-optic modulator. During some locks, these 
radio frequency beat notes were extremely problematic and were responsible for 
$\sim$90\% of all glitches seen in DARM at LLO (Figure \ref{hveto_llo}). When 
testing the binary black hole search on Livingston aLIGO data containing these 
radio frequency beat note glitches, they proved to greatly limit the search 
performance. 

\begin{figure}
\centering
\subfloat[Before\label{before_whistles}]{\includegraphics[width=0.5\textwidth]{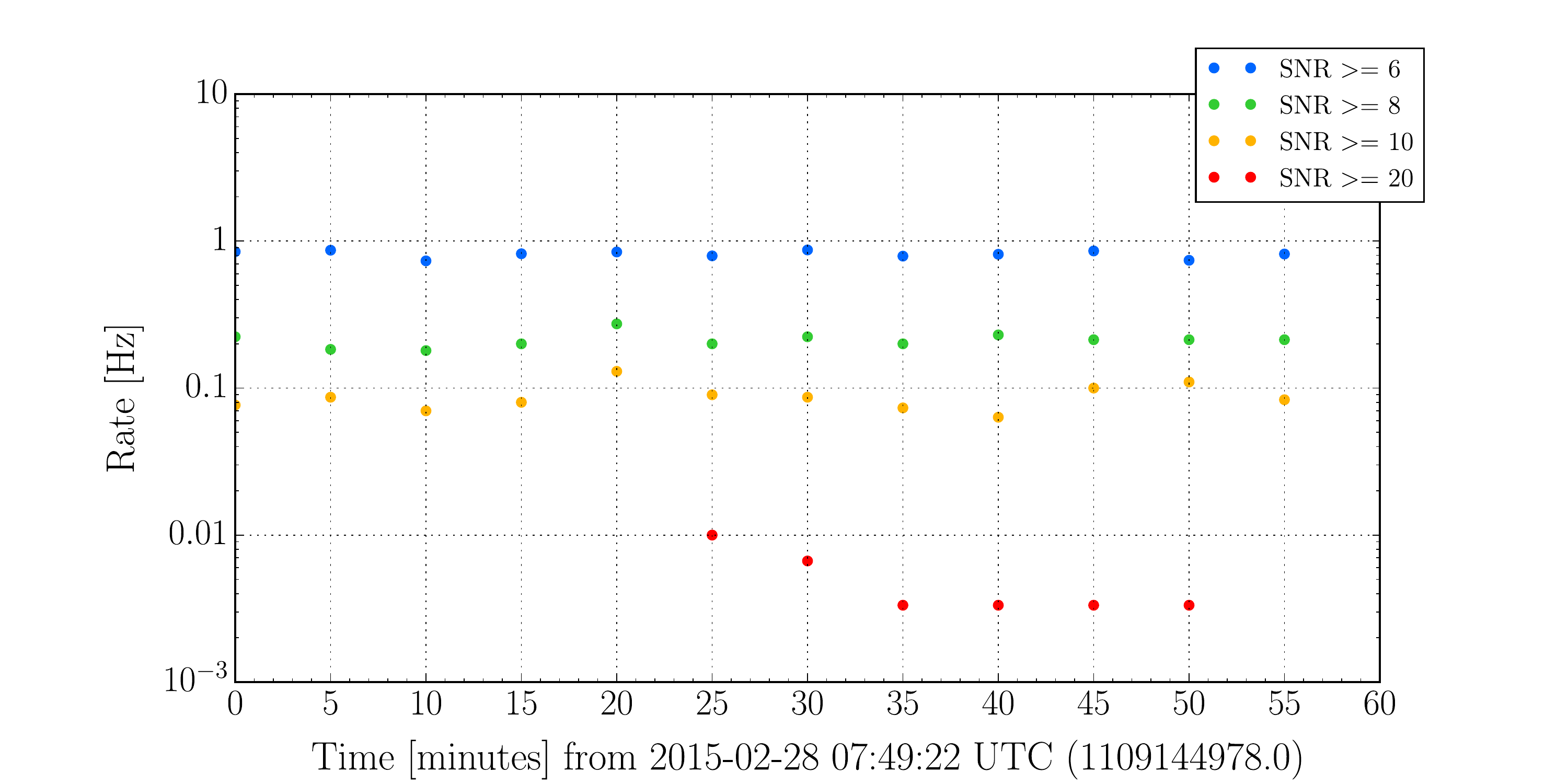}}
\subfloat[After\label{after_whistles}]{\includegraphics[width=0.5\textwidth]{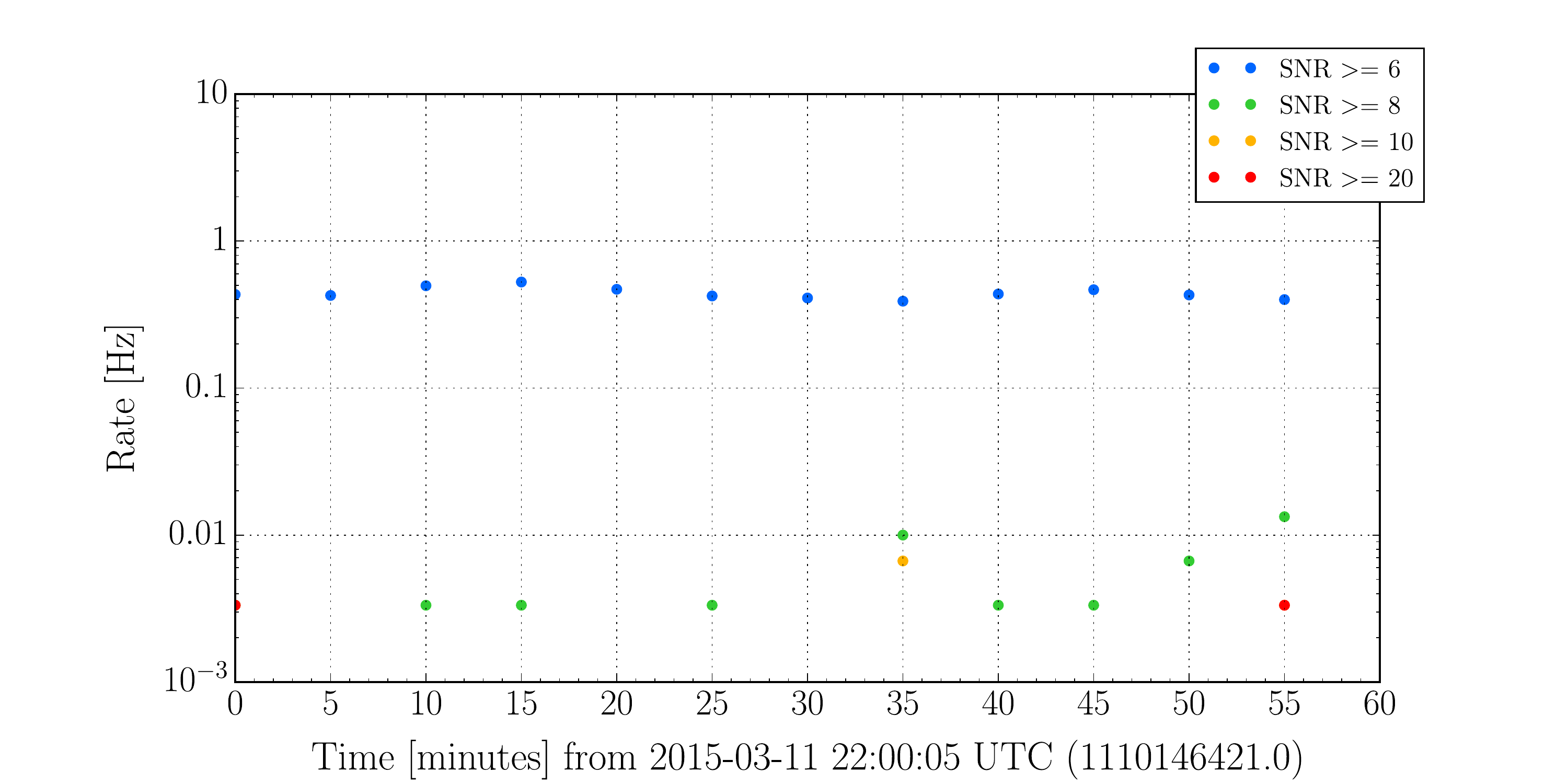}}
\caption{\label{gr_arches_llo}The rate of Omicron triggers before (Figure 
\ref{before_whistles}) and after (Figure \ref{after_whistles}) the 
problematic voltage controlled oscillator was moved at Livingston. The 
plots illustrate an hour of data and the triggers are separated in to SNR 
bins where the rate is calculated over a five minute period.}
\end{figure}

To remove these sources of glitches from DARM, the output frequencies of the 
problematic voltage controlled oscillators were moved away from each other in
frequency. Once this had been accomplished, a dramatic decrease in the number 
of glitches associated to these sources was observed. At LLO, the 
overall glitch rate improved by a factor of two and glitches with an SNR $>$ 8 
improved in rate by a factor of 50, as shown in Figure \ref{gr_arches_llo}. 
However there are numerous voltage controlled oscillators with output 
frequencies in the same vicinity, meaning radio frequency beat notes could 
and have presented themselves again as commissioning continues. Due to the 
work presented, there are now more deterministic methods for finding and 
tracking these types of glitches.

\subsection{Pre-Stablized Laser Periscope}
The pre-stabilized laser subsystem consists of a Nd:YAG laser and control 
systems which stabilize the laser in beam direction, frequency, and intensity 
\cite{Kwee:2012kw}. The system lives outside of the vacuum system (which 
houses the majority of the aLIGO hardware) in a clean, dust-free environment. 
The optical components of the pre-stabilized laser are located on an optical 
bench, which is, 
physically, much lower than the height the beam needs to be to enter the 
vacuum systems viewport. A periscope is used to elevate the beam so 
that it can enter the vacuum enclosure. 

\begin{figure}
\centering
\includegraphics[width=\textwidth]{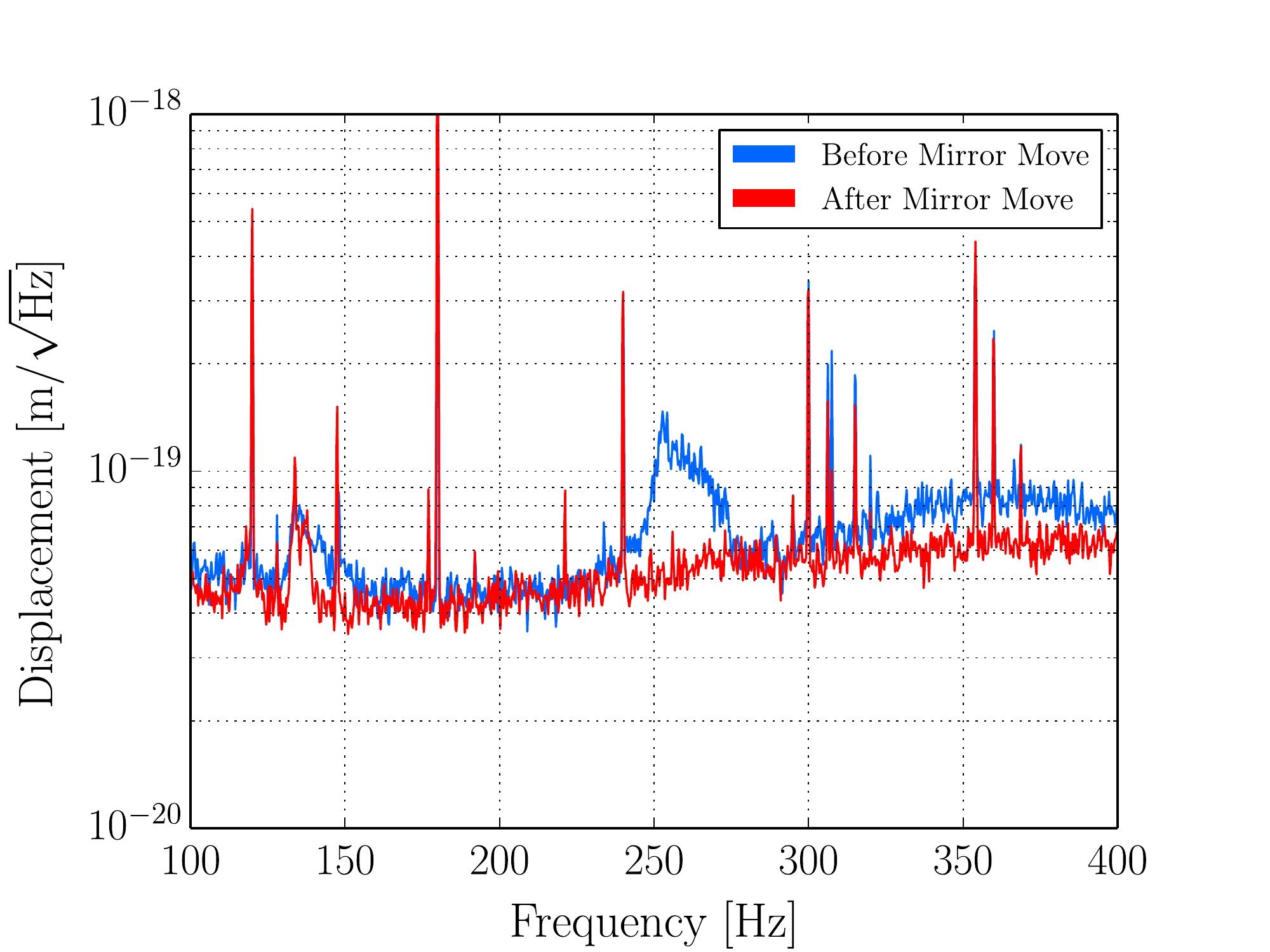}
\caption{\label{pztmove}Spectrum of DARM before and after the piezo-actuated 
mirror was relocated from the top of the pre-stabilized laser periscope 
to the optical bench. When the mirror was located at the top of the 
periscope, excess beam jitter can be seen at 135 and 250 Hz. After the move, 
the former feature is reduced and the latter feature is completely removed.}
\end{figure}

Input pointing noise can couple to DARM through several mechanisms. As 
shown in Figure \ref{aligo_layout}, the laser beam propagates inside the input 
mode cleaner (IMC) before entering the main interferometer. Any jitter at the 
input of the IMC cavity is filtered, since the cavity is stable. However, 
jitter is also converted into intensity noise in transmission through a 
quadratic coupling, which can have a linear component if the IMC cavity is 
slightly misaligned. Intensity noise then couples to DARM since the Advanced 
LIGO detectors are using a DC readout scheme \cite{Fricke:2012dc} in which the 
primary laser is sensed 
on a DC photodiode and used as a reference to identify gravitational wave 
sidebands. 
Clearly, one solution is to tune 
the IMC angular controls to improve its alignment. However, the residual 
misalignment was large enough to still introduce a non negligible amount of 
intensity noise at the input of the main interferometer. 
For example, two distinct features in the DARM spectrum at 
$\sim$135 Hz and $\sim$250 Hz were due to the resonance of a mirror mount securing 
a piezo-actuated mirror, used to control the beam pointing in to the input mode cleaner. 
This mirror mount was located at the top of the 
pre-stabilized laser periscope and was causing excess beam jitter. 
To relieve this issue, the piezo-actuated mirror was moved from 
the top of the pre-stabilized laser periscope to the pre-stabilized laser 
optical table and was replaced with a 
fixed mirror. This immediately improved the input pointing noise from the 
PSL to the vacuum enclosure, reducing the effect at 
135 Hz and removing the 250 Hz resonance, as seen in Figure \ref{pztmove}.

\subsection{End Test Mass Ring Heaters}
The laser power circulating in the aLIGO arms can be as large as 750 kW, which 
will cause deformations in the test masses via thermal expansion, changing the 
spatial modes of the 
optical field \cite{Aasi:2015fr}. Thermal compensation techniques are used to 
maintain the radius of curvature of the input and end test masses, 
which, if left uncompensated, would increase by a few tens of meters 
\cite{Aasi:2015fr}. A ring heater is one of the elements of the thermal 
compensation system used in aLIGO. It comprises a nickel-chromium heater wire 
wound 
around two semi-circular glass rods and radiates onto the test mass to 
compensate for thermo-elastic deformations of the test masses 
\cite{Aasi:2015fr}.

\begin{figure}
\centering
\includegraphics[width=\textwidth]{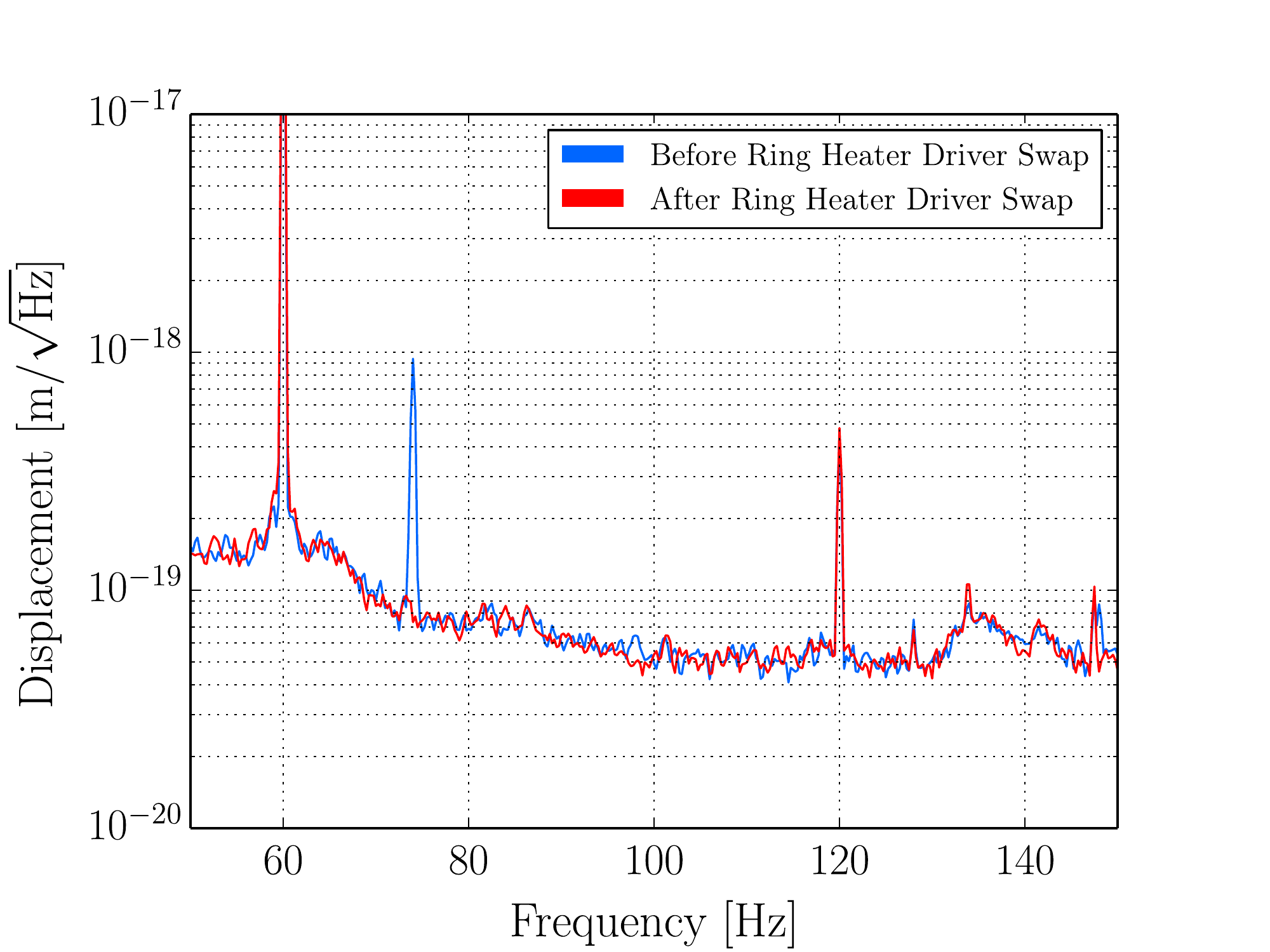}
\caption{\label{spectrum_ringheater}Spectrum of DARM before and after the ring 
heater at the end X station was replaced. Notice the removal of the 76 Hz 
spike in the spectrum once the ring heater driver is replaced. The 60/120 Hz 
features are due to the mains power.}
\end{figure}

Because ring heaters actuate on the major optics of the 
interferometer, noise associated with them can couple significantly into DARM. 
This was the case for a feature found during commissioning activities at 
Livingston, at 74 (end X station) and 76 Hz (end Y station) in the DARM 
spectrum. This noise feature 
was found to be coherent with a magnetometer at the same end station. 
Upon further investigations, it was discovered the ring heater drivers were 
the source of the added noise. Subsequently, they were both replaced 
with drivers modified to have filtered power supplies. The improvement in the 
gravitational wave spectrum for the replacement at the Y end can be seen in 
Figure \ref{spectrum_ringheater}. Similar features in DARM due to the 
ring heaters were also seen at Hanford. This discovery and increased 
understanding of the potential systematic impacts of the thermal compensation 
system has prompted the setup of monitors which will catch such future problems.

\subsection{LLO Pre-Mode Cleaner}
The pre-mode cleaner (PMC) is a four mirror resonator in a bow tie 
configuration responsible for removing higher-order modes from the beam,
reducing beam jitter, and providing low-pass filtering for radio 
frequency intensity fluctuations \cite{Aasi:2015fr}. Three mirrors are fixed, 
while the fourth is attached to a piezoelectric transducer (PZT) to control 
the length of the pre-mode cleaner. In terms of the interferometer layout 
the pre-mode cleaner is located just before the input mode cleaner (see 
Figure \ref{aligo_layout})

\begin{figure}
\centering
\includegraphics[width=\textwidth]{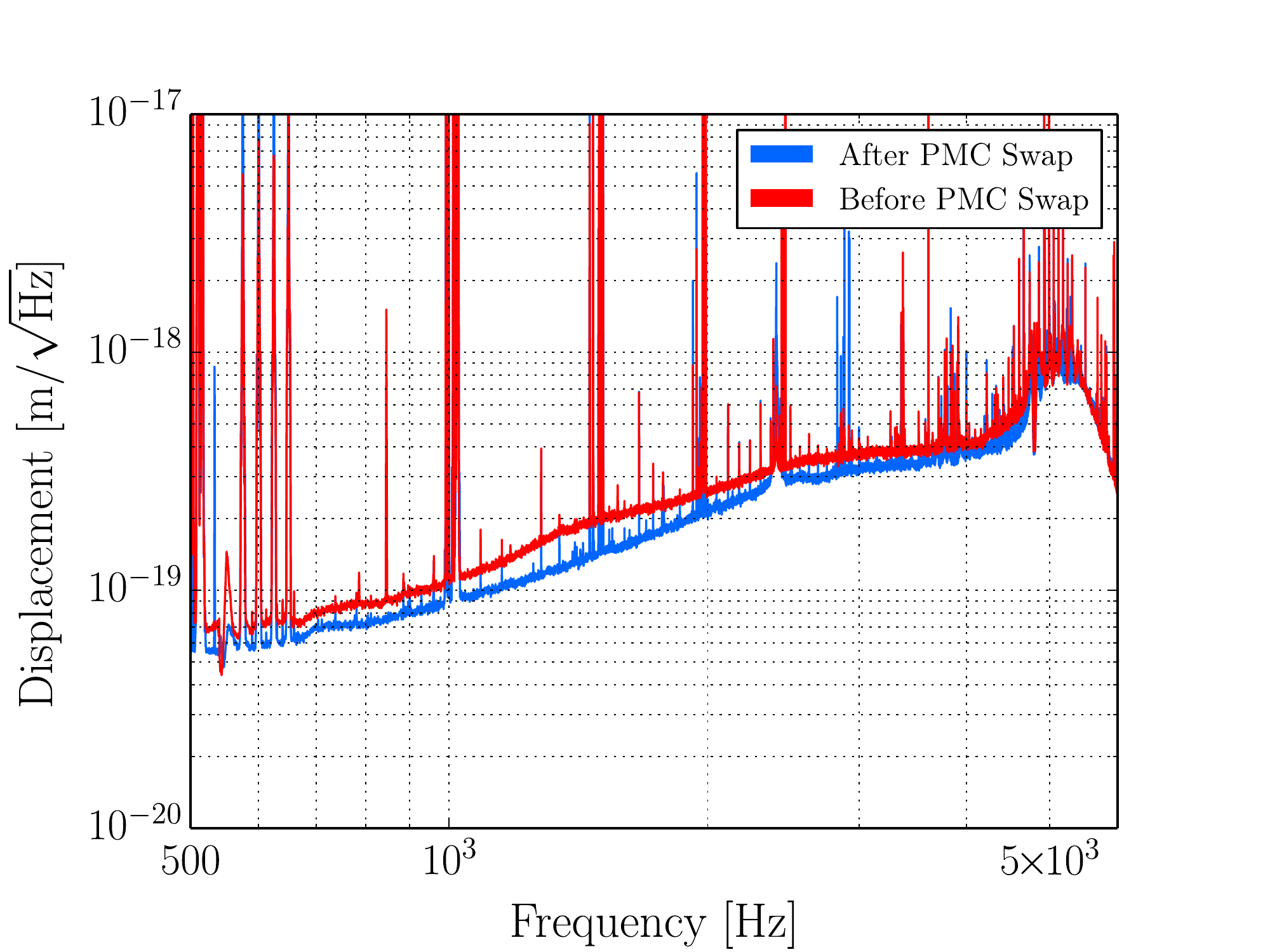}
\caption{\label{spectrum_pmcnoise}The DARM spectrum before and after the 
pre-mode cleaner was swapped. The elevated noise between 1-2 kHz is 
reduced with the change.}
\includegraphics[width=\textwidth]{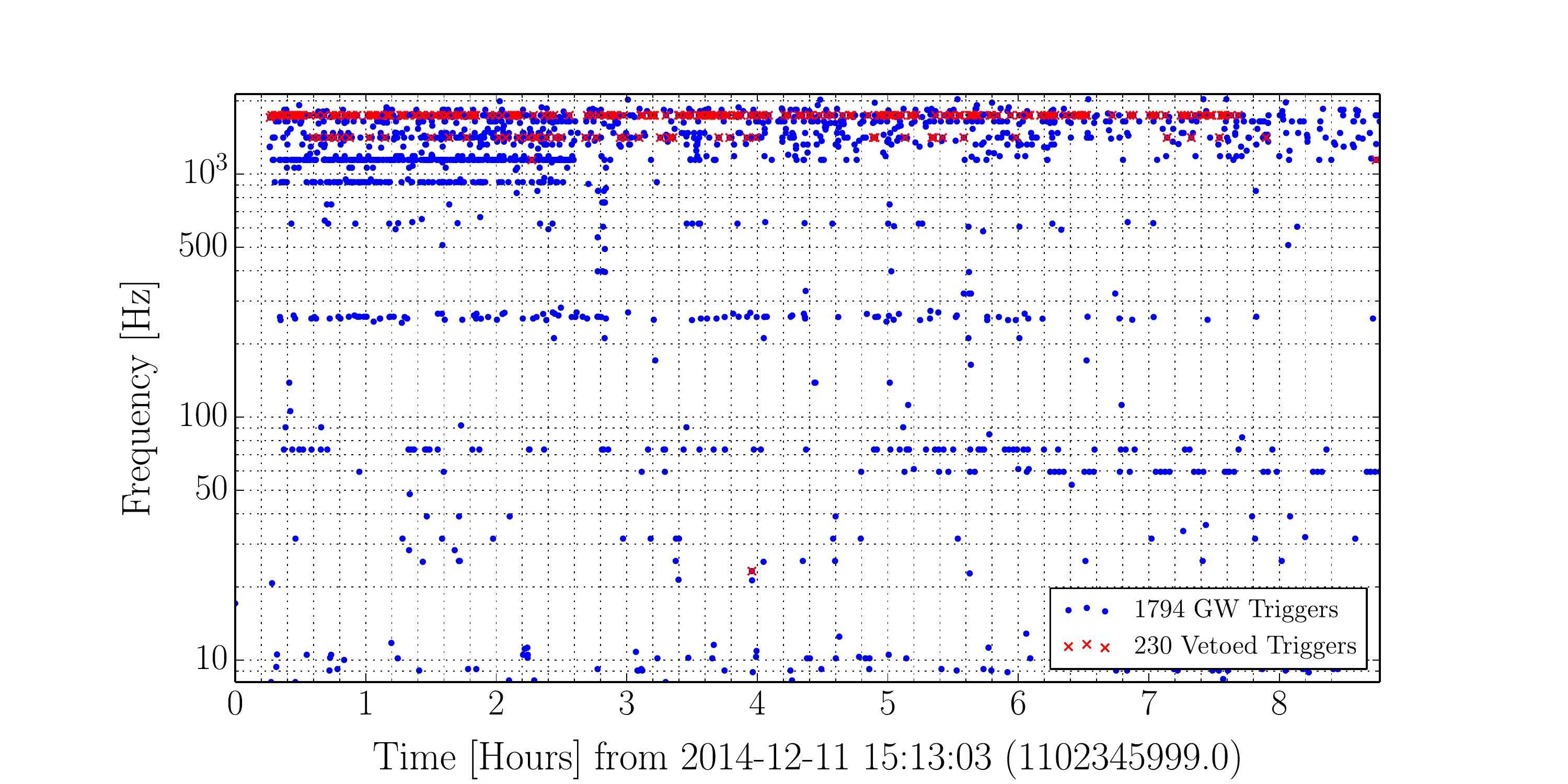}
\caption{\label{hveto_pmc}Nine hours of noise events recorded by Omicron. 
The blue points show all of the triggers in DARM while the red `x' overlay 
designates those which are identified by HVeto as being coincident with glitches
in the the pre-mode cleaner drive. The coincident triggers are mainly above 1 kHz in DARM.}
\end{figure}

During locking activities in early December 2014, noise between $1 - 2$ kHz 
(see Figure \ref{spectrum_pmcnoise}) in DARM was found to be correlated with 
the pre-mode cleaner system. Not only was there found to be a linear 
coherence between DARM and the pre-mode cleaner in this frequency range, Hveto 
also found the glitches in the 
high voltage drive that controls the PZT, which is fixed to one of the 
four pre-mode cleaner mirrors, to be coincident with glitches seen 
between $1 - 2$ kHz in DARM. Figure \ref{hveto_pmc} shows the statistical 
correlation seen by HVeto.
The decision was made to completely replace the pre-mode cleaner. Before the 
swap, however, it was discovered that cables left over from previous 
measurements were connected to electronics used to interact with the PMC. 
These additional cables were demonstrated to cause the increased noise in 
the DARM channel. Due to the sensitive nature of the LIGO instruments, subtle 
details like improperly terminated cables can have a serious impact on data 
quality, and people both on and off site are constantly monitoring to catch 
these kinds of problems. Although this additional noise in DARM was not found 
to be due solely to pre-mode cleaner degradation, the pre-mode cleaner was 
still replaced to improve other areas of operation.

\section{Comparing the Initial and Advanced LIGO Performance of the LIGO Livingston Detector}\label{sec:comparison}
%During LIGO's sixth science run some of the best data was ever collected in
%the search for gravitational waves. Although a detection was not made, many
%astrophysically interesting statements could be made (for example
%\cite{Abadie:2012cb}, \cite{Abadie:2012bu}, \cite{Aasi:2014pu}). However the
The non-Gaussian, non-stationary nature of the noise in the detectors has a
great impact on the ability to perform a search for gravitational waves.
The glitches and features described in Section \ref{sec:DQ} would have a 
detrimental impact
on the transient analyses had they not been removed. Loud glitches (in terms of
SNR) can mask or potentially disrupt a transient gravitational wave signal in
the data, whereas many quieter glitches can raise the background of a search 
and limit the ability to distinguish signal from noise.

Section \ref{sec:comp_aLIGO} looks at the aLIGO Livingston detector before and 
after the DQ issues discussed in Section \ref{sec:DQ} were resolved. Section 
\ref{sec:comp_S6} compares Livingston's S6 detector output to the early aLIGO 
data after the removal of the example DQ issues. For both cases, six hours of 
detector data were chosen to 
best represents the detectors at the time of operation. The S6 data 
were taken from September 16th 2010, which is considered to be a typical, 
good, and stable stretch of data. All subsequent plots showing data from 
this time are labelled `S6'. Data from December 16th 2014, which were 
taken during an engineering run, contained all the DQ issues 
previously described and are labelled as `ER6' in the subsequent plots in this 
section. Data from a lock stretch on March 11th 2015 were selected to represent 
the early aLIGO data (and is labelled `Post ER6' in the following plots). 
In terms of data quality and inspiral range, the March 11th aLIGO 
data represents some of the best from that period. However, it should be noted 
that 
this latter stretch of data represents active commissioning time, where the 
detector is not left completely alone and is therefore not as stable as the S6 
or ER6 data. Since the writing of this paper it is worth noting that another 
engineering run has taken place.

\subsection{A Comparison of aLIGO Data, Before and After DQ Example Issues 
were Resolved}\label{sec:comp_aLIGO}
For two weeks in December 2014, the LIGO Livingston detector took part in an 
engineering run (ER6). During this time, the detector was locked as 
often as possible, with minimal interruption for commissioning activities, to 
assess the stability and quality of the data. 

\begin{figure}
\centering
\subfloat[ER6 vs Post ER6 Trigger Rate Histogram]{\includegraphics[width=0.5\textwidth]{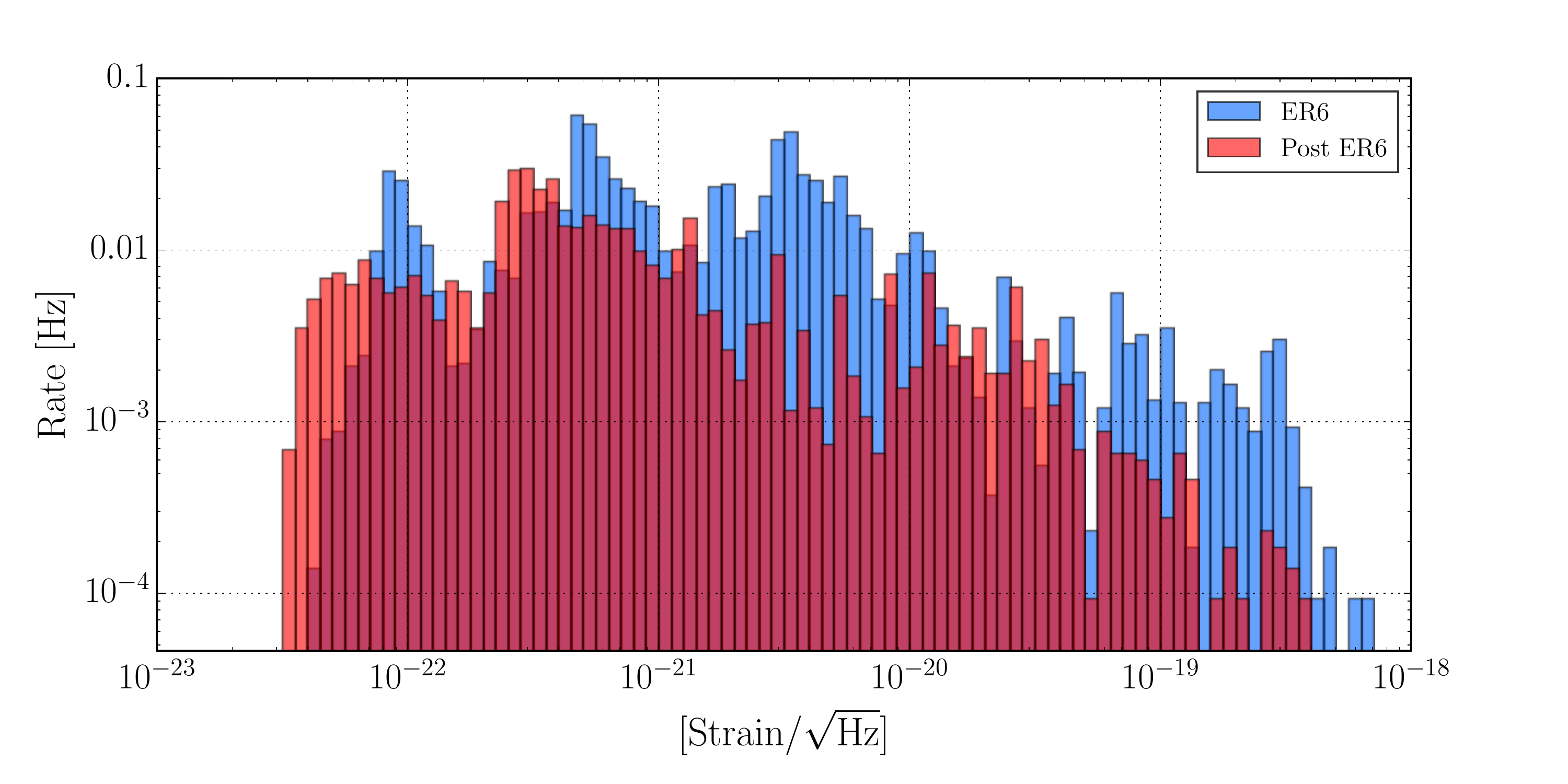}\label{ER6_trh}}
\subfloat[ER6 vs Post ER6 Trigger Spectrum]{\includegraphics[width=0.5\textwidth]{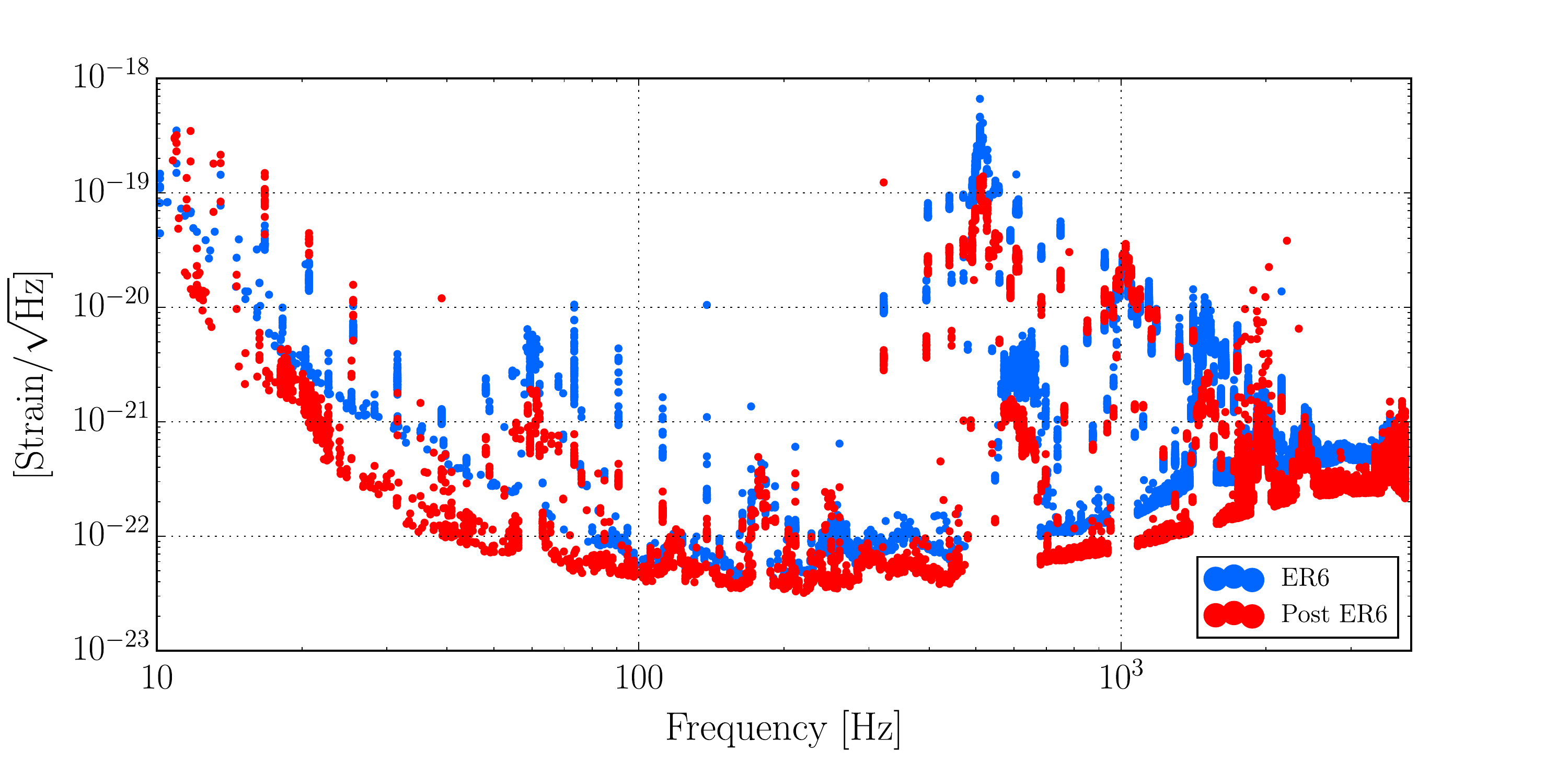}\label{ER6_ts}}\\
\subfloat[ER6 Trigger Rate]{\includegraphics[width=0.5\textwidth]{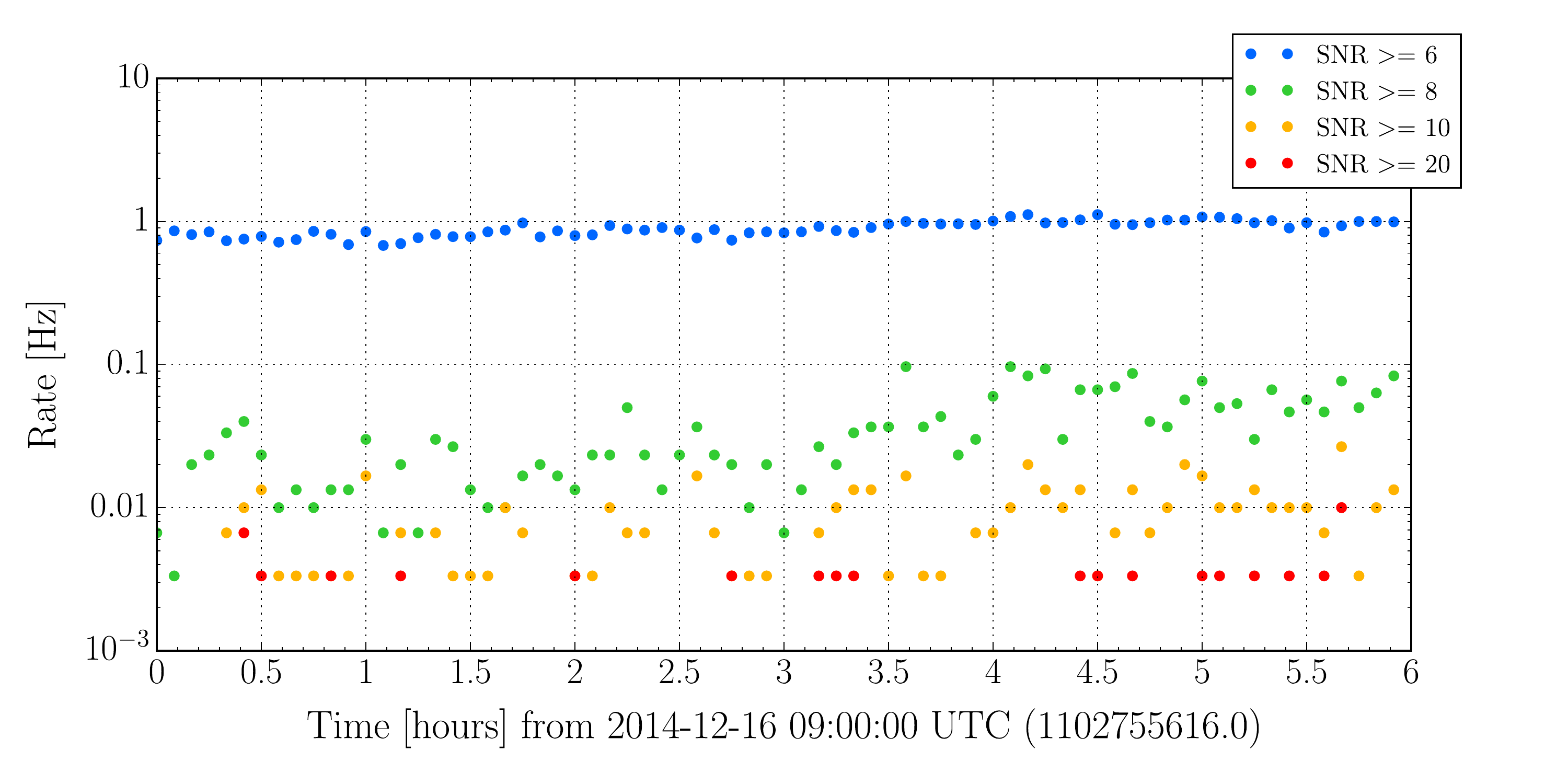}\label{ER6_tr}}
\subfloat[Post ER6 Trigger Rate]{\includegraphics[width=0.5\textwidth]{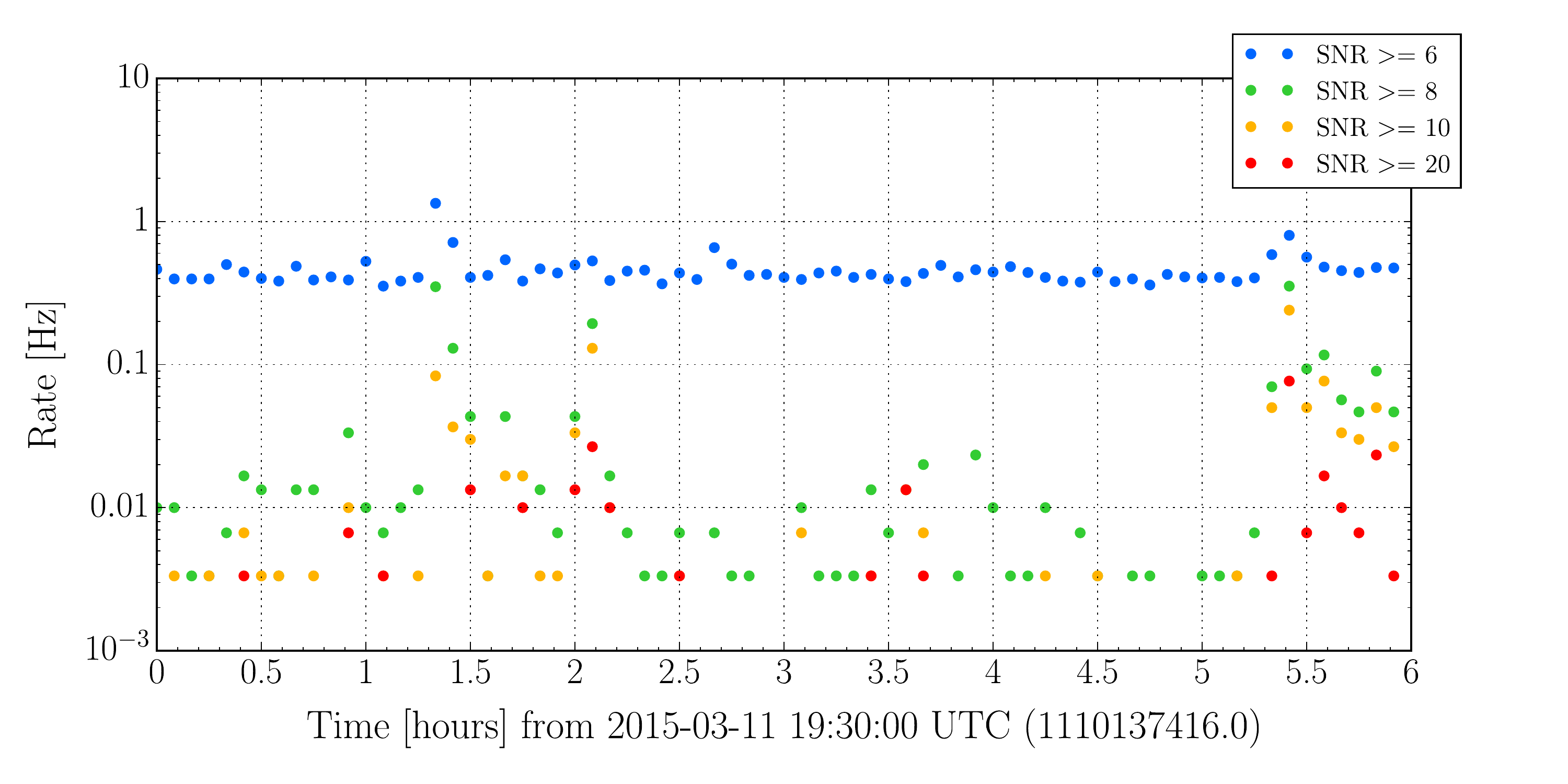}\label{ER6_aLIGO_tr}}\\
\subfloat[Gaussian Noise Trigger Rate]{\includegraphics[width=0.5\textwidth]{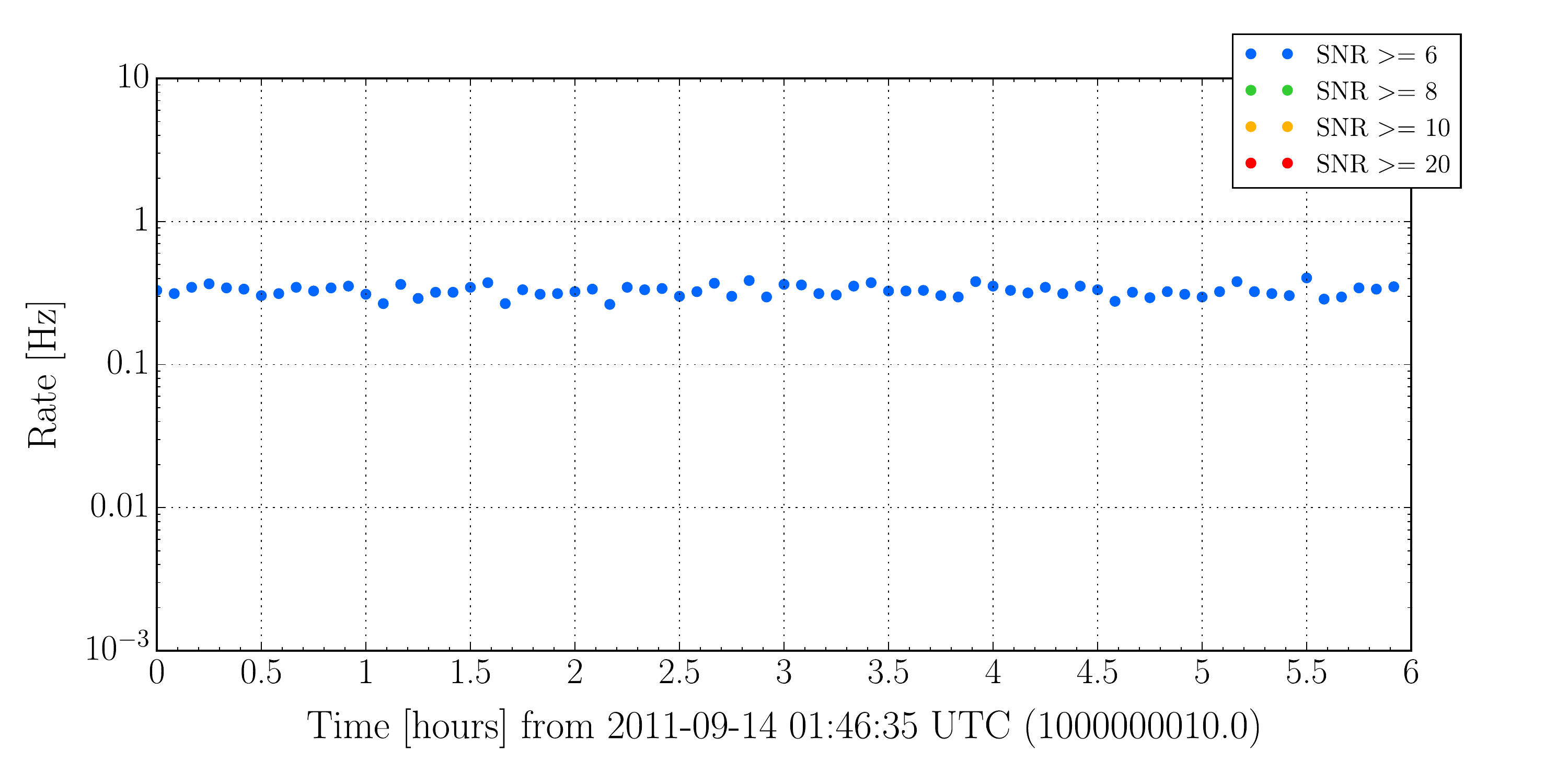}\label{design_tr}}
\caption{\label{comparison_ER6}
A comparison of ER6 and Post ER6 data. Figure 
\ref{ER6_trh} shows a trigger rate histogram, 
binned by strain/$\sqrt{\mathrm{Hz}}$. Figure \ref{ER6_ts} is the trigger 
spectrum between the two 
times, which gives a sense of the most `glitchy' frequencies. Figures 
\ref{ER6_tr} and \ref{ER6_aLIGO_tr} are the trigger rate plots for both the 
ER6 and Post ER6 time period respectively. In these plots the Omicron triggers 
are separated in to four SNR bins and the data is combined over a five minute 
interval to calculate the rate. The first four plots show the improvement in 
the data quality, as witnessed by Omicron, between ER6 and the Post ER6 time 
period chosen. For comparison, Figure \ref{design_tr} shows the trigger rate 
for Gaussian noise at design sensitivity.}
\end{figure}

Figure \ref{comparison_ER6} compares the data quality 
between ER6 and the Post ER6 data from the viewpoint of Omicron triggers. 
Figure \ref{ER6_trh} is a trigger rate histogram, which compares the rate of 
Omicron triggers produced over the two data, binned by amplitude spectral 
density in strain/$\sqrt{\mathrm{Hz}}$. From this figure two things are 
evident. First, the sensitivity of the detector has improved, resulting in 
strain/$\sqrt{\mathrm{Hz}}$ bins that extend to lower values. Second, the rate 
of glitches at all amplitudes is in general lower. Figure \ref{ER6_ts} 
shows the Omicron trigger spectrum for both ER6 and Post ER6. This figure gives 
a sense of which frequencies are contributing the most to a certain 
strain/$\sqrt{\mathrm{Hz}}$ 
bin from the previous figure. At $\sim$500 Hz (and harmonics) are 
the test mass suspensions violin modes \cite{Aasi:2015fr}, which seem to be 
the worst offenders in terms of amplitude.
Damping these violin modes further would presumably decrease the rate of 
higher amplitude glitches seen in the data. Between the two times 
analysed, there has been significant improvement. Figure \ref{ER6_tr} shows the 
trigger rate for the ER6 data, binned (colored) by SNR over five minute 
segments. Figure \ref{ER6_aLIGO_tr} displays the same information for the 
Post ER6 data. Despite the stability of the Post ER6 data varying more than the 
ER6 data (there are disturbances which are causing the glitch rate to 
increase in the March 2015 lock at $\sim$1.5 and $\sim$5.5 hours), the rate of 
triggers for the data seems to decrease. The rate of low SNR glitches (SNR of 
5) has halved between the two times, and, ignoring the increased glitch rate 
at the two highlighted times, the rate of glitches with an SNR $\geq$ 8 and 10 
has decreased by around an order of magnitude. For comparison Figure 
\ref{design_tr} shows the expected trigger rate for Gaussian noise with a 
sensitivity equivalent to that at design (circa 2018). This figure represents 
the best the data could be.

\subsection{A Comparison of S6 and Early aLIGO Data}\label{sec:comp_S6}
During initial LIGO's sixth science run (S6), some of the best quality 
observing run data was collected. Although a detection 
was not made, many astrophysically interesting statements were made from the 
searches conducted (for example
\cite{Abadie:2012cb}, \cite{Abadie:2012bu}, \cite{Aasi:2014pu}). Many people, 
both on and off site, have been working to improve the output of the detectors, 
not only to increase the volume over which to search for gravitational waves, 
but also to provide data which are as stationary and Gaussian as possible.

\begin{figure}
\centering
\subfloat[S6 vs Post ER6 Trigger Rate Histogram]{\includegraphics[width=0.5\textwidth]{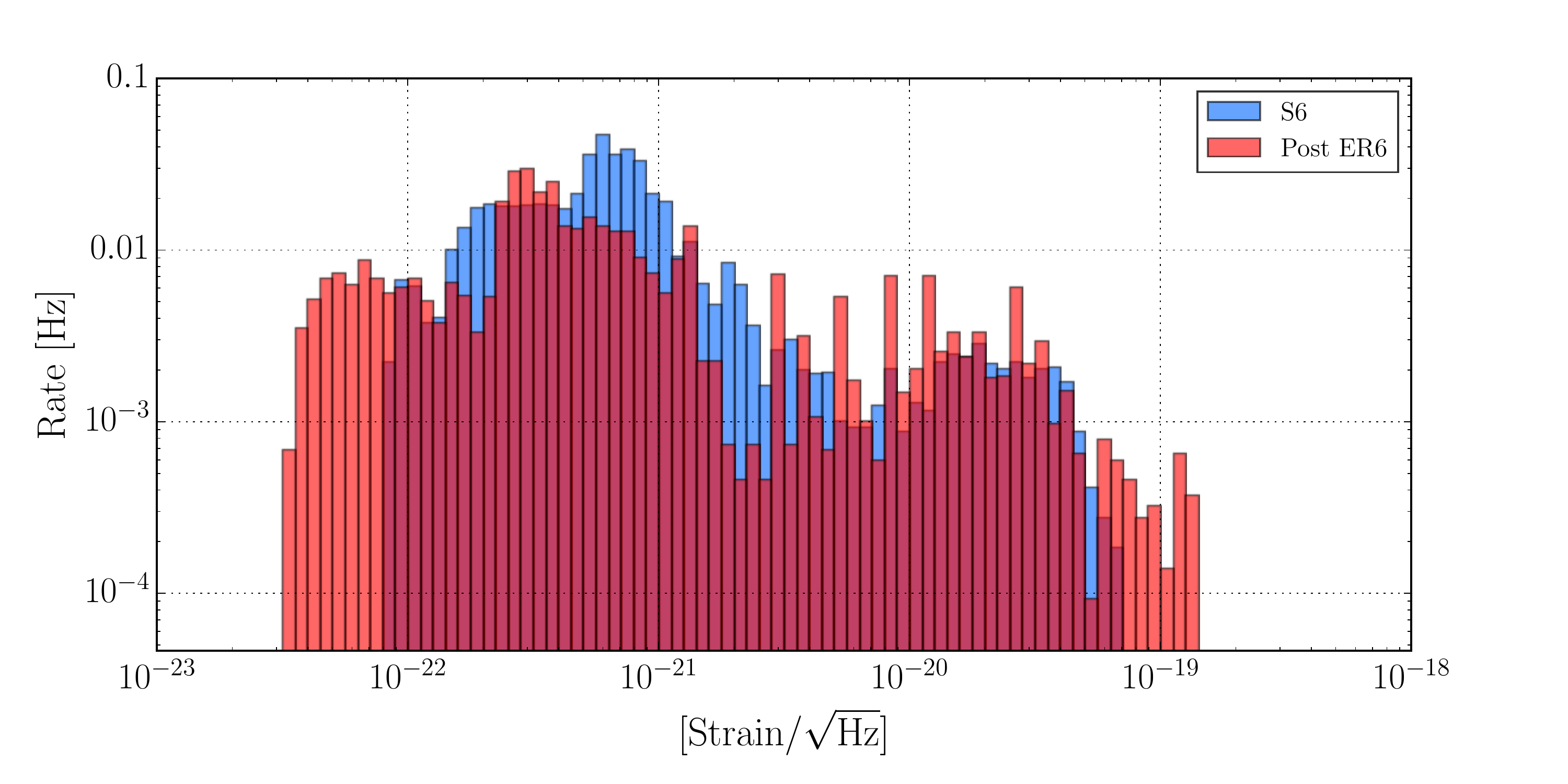}\label{rate_strain}}
\subfloat[S6 vs Post ER6 Trigger Spectrum]{\includegraphics[width=0.5\textwidth]{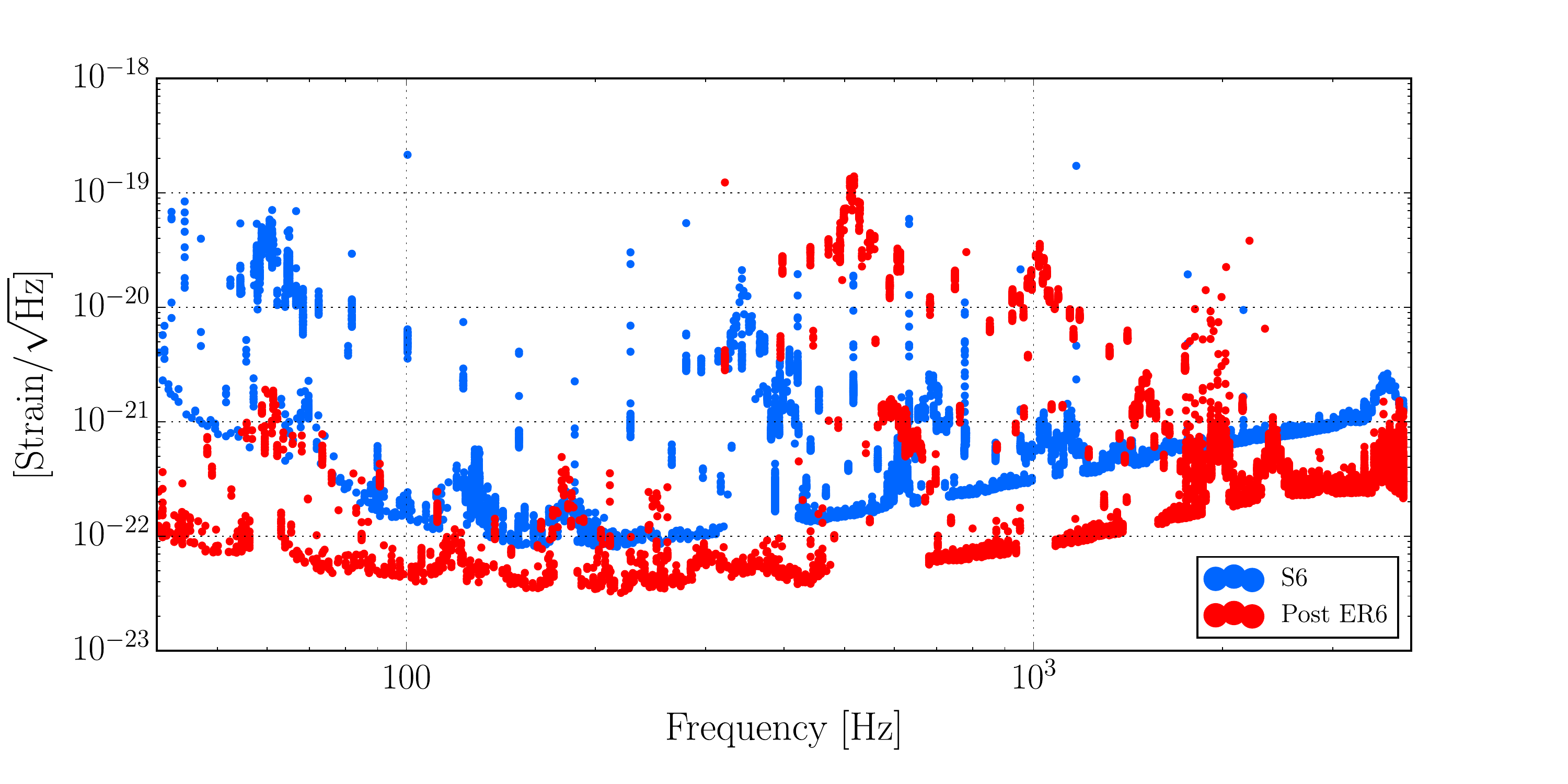}\label{trig_spec}}\\
\subfloat[S6 Trigger Rate]{\includegraphics[width=0.5\textwidth]{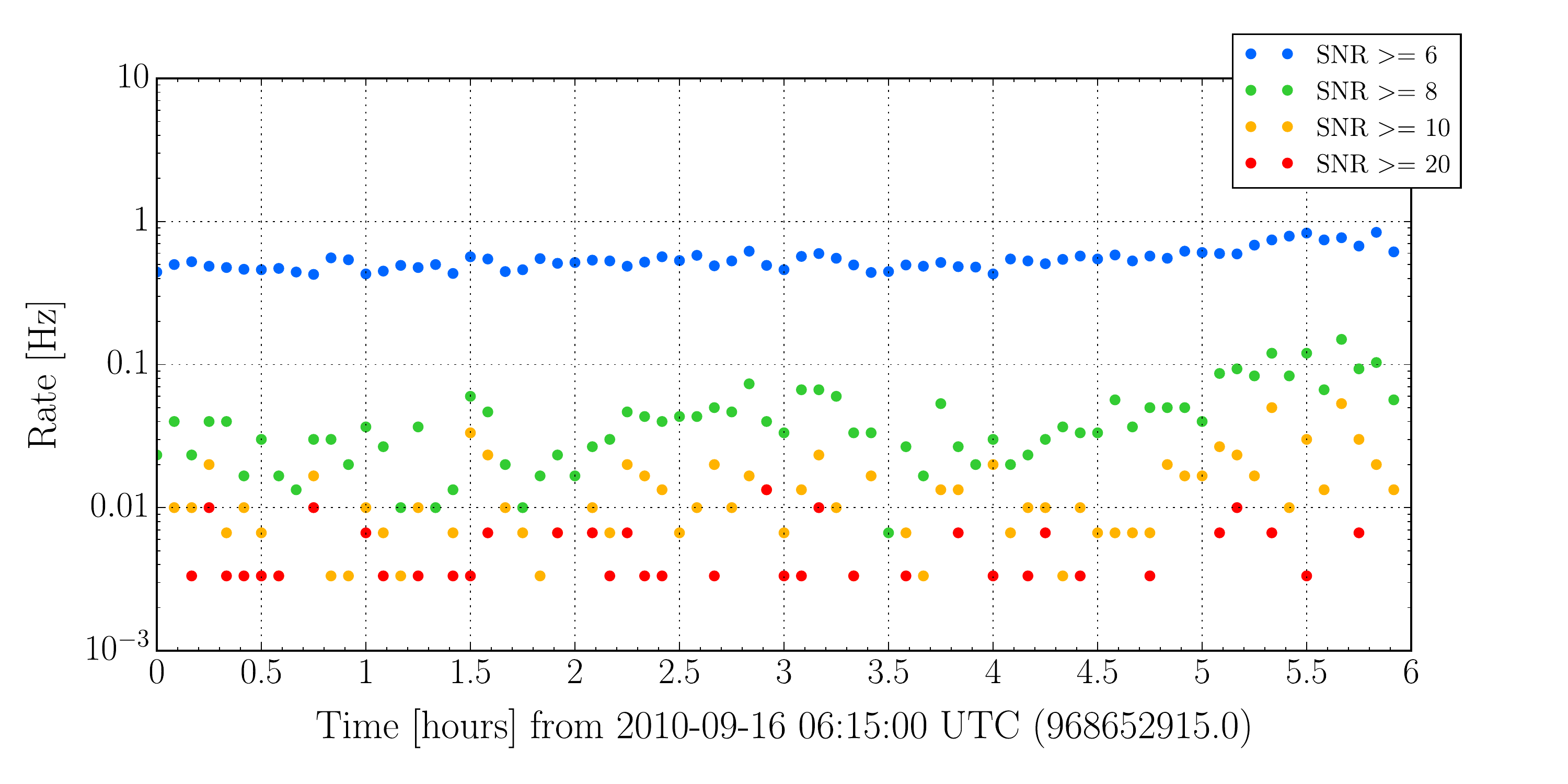}\label{rate_s6}}
\subfloat[Post ER6 Trigger Rate]{\includegraphics[width=0.5\textwidth]{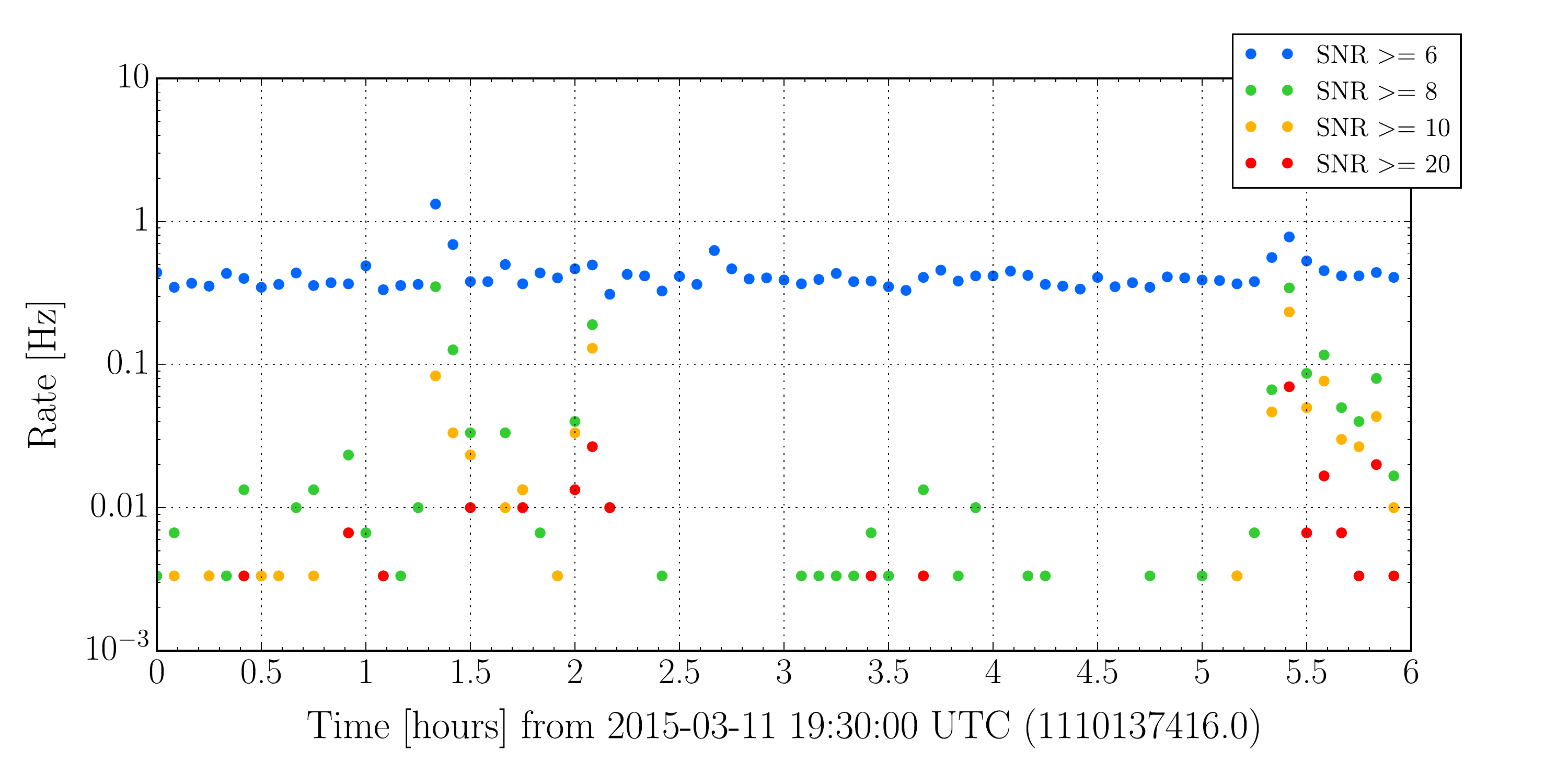}\label{rate_aLIGO}}
\caption{\label{comparison}Comparison plots of six hours of data taken in S6 
and the time period chosen to represent early aLIGO data. Unlike 
Figure \ref{comparison_ER6} these plots are in the frequency range 40-4096 Hz, 
as during S6 40 Hz was the lowest frequency we searched for gravitational 
waves. Figure \ref{rate_strain} shows a trigger rate histogram, in terms of 
strain and Figure \ref{trig_spec} is the trigger spectrum between S6 and Post 
ER6. 
Figures \ref{rate_s6} and \ref{rate_aLIGO} are the trigger rate plots for both 
the S6 and Post ER6 time period respectively. 
The Omicron triggers are separated 
in to four SNR bins and the data is combined over a five minute for these last 
two plots.}
\end{figure}

Figure \ref{comparison} shows a comparison of LIGO Livingston's 
detector data from S6 and from a lock stretch in March 2015 (Post ER6). Figure 
\ref{rate_strain} compares the Omicron trigger rates binned by 
strain/$\sqrt{\mathrm{Hz}}$. It is clear to see that the triggers produced 
using 
Post ER6 data have a lower floor in strain/$\sqrt{\mathrm{Hz}}$ than in S6 
($<10^{-22}$) and the rate of glitches is generally lower across the entire 
amplitude range. The higher 
amplitude glitches ($\sim10^{-19}$) in the aLIGO data seem to be 
caused by the band of glitches around 500 and 1000 Hz, depicted in figure 
\ref{trig_spec}. These are the fundamental and first harmonic of the test mass 
violin modes (as previously mentioned). For S6, these modes were closer to 340 
and 680 Hz \cite{Abbott:2009ab}. It is worth noting that the violin modes in 
aLIGO have a much higher Q than in Initial LIGO, hence they are more easily 
excited to super-thermal levels \cite{Aasi:2015fr}. 
Figure \ref{trig_spec} compares the triggers seen in both S6 and Post ER6 in 
terms of their frequency and amplitude. This plot 
highlights some of the most problematic regions to a transient search, in the 
frequency spectrum. Figures \ref{rate_s6} and \ref{rate_aLIGO} show the rate of 
triggers, separated into four SNR bins over five minute intervals for the S6 
and Post ER6 data respectively. 
Overall, the rate of triggers was less in the aLIGO March data 
compared to S6; the real difference can be seen in the rate of higher SNR 
triggers. Despite the variability of the aLIGO data (instrumental 
features at $\sim$1.5 and 5.5 hours in Figure \ref{rate_aLIGO} cause the 
glitch rate to increase), the glitch rate in certain SNR bins (SNR 8 - 20) is 
half to an order of magnitude lower in the March 2015 data than it was in 2010 
(i.e. S6).

% Summary or maybe concluding remarks
\section{Concluding remarks}\label{sec:conclusion}
%instrumental fixes helped - see plots
%better shape than s6
%aLIGO is ready
The Advanced LIGO detectors have been completely assembled and are currently 
being optimized to begin operation later this year, in what will be the most 
sensitive search for gravitational waves to date. In addition to maximizing 
the inspiral range, and therefore the volume over which an astrophysical search 
is conducted, it is just as important to ensure the data are as Gaussian and 
stationary as possible. In the last science run of initial LIGO (S6), many 
issues were 
identified as detrimental to a search for gravitational waves and harmful 
to the stability of the interferometers. Improvements were made in that 
some sources of noise were fixed at the instrument, but DQ flags were 
often used to identify known correlations between noise in an auxiliary 
channel and the gravitational wave channel \cite{Aasi:2015de}. For aLIGO, the 
hope is to identify and resolve harmful sources of noise as much as possible 
before an analysis is performed.

There have been many improvements to the aLIGO detectors during their 
commissioning phase; many issues have been identified and resolved at the 
source and this work continues. 
Section \ref{sec:comp_aLIGO} highlights the improved 
quality of the detector data during this commissioning period, with the 
rate of glitches decreasing over the period studied. These same data were
also compared with those from the last science run. The rate of 
glitches is less now than that seen in the last science run (Section 
\ref{sec:comp_S6}). This work focused on data from the LIGO-Livingston 
detector and similar work is underway at LIGO-Hanford which will be 
presented in a future publication.

Over the coming months more 
effort will be made to further improve the output of the detectors to 
provide the best possible quality data to search for gravitational waves. 
Early efforts have already proven to be vital.

\section{Acknowledgements}
The authors would like to thank Gabriela Gonz\'{a}lez, Peter Saulson and 
David Shoemaker for useful discussions throughout the writing of this paper. 
LKN was supported by the College of Arts and Sciences at Syracuse University, 
TJM and RPF were supported by NSF award PHY-1205835 and DMM by NSF award 
PHY-1104371. JRS was supported by NSF CAREER 1255650. ARW was supported by the 
UK Science and Technology Facilities Council through grants ST/K501931/1 and 
ST/L000962/1. Some calculations were performed on the ORCA cluster supported 
by NSF award PHY-1429873.
LIGO was constructed by the California Institute of Technology and 
Massachusetts Institute of Technology with funding from the National Science 
Foundation, and operates under cooperative agreement PHY-0757058. Advanced LIGO 
was built under award PHY-0823459. This paper has the LIGO Document Number 
LIGO-P1500131.

\section{References}
\bibliographystyle{unsrt}
\bibliography{references.bib}

\end{document}